\documentclass[letterpaper,twocolumn,10pt]{article}
\usepackage{usenix} 
\pdfoutput=1
\usepackage[T1]{fontenc}
\usepackage[utf8]{inputenc}
\usepackage{microtype}
\microtypesetup{final}
\UseMicrotypeSet[protrusion]{basicmath}

\usepackage{amsmath,amssymb,amsfonts,amsthm}
\usepackage{authblk}
\usepackage{mathtools}
\usepackage{bm}
\usepackage{booktabs}
\usepackage{multirow}
\usepackage{xcolor}
\usepackage{graphicx}
\usepackage{hyperref}
\usepackage{algorithm}
\usepackage{algpseudocode}
\usepackage{enumitem}
\usepackage{cleveref}
\usepackage{url}
\usepackage{listings}
\usepackage{float} 
\usepackage{tabularx}
\usepackage{graphicx}
\usepackage{subcaption}
\usepackage{ragged2e}
\usepackage{etoolbox}
\AtBeginEnvironment{algorithmic}{\RaggedRight}

\usepackage{booktabs}
\usepackage{siunitx}
\usepackage{array}
\usepackage{colortbl}
\usepackage{xcolor}
\usepackage{circledsteps}

\usepackage{booktabs}
\usepackage{siunitx}
\usepackage{array}
\usepackage{colortbl}
\usepackage{xcolor}
\usepackage{tabularx}
\usepackage{booktabs}
\usepackage{tabularx}
\usepackage{siunitx}
\usepackage{xcolor,colortbl}
\hypersetup{
  colorlinks=true,
  linkcolor=blue!55!black,
  citecolor=blue!55!black,
  urlcolor=blue!55!black
}

\let\VAorigtexttt\texttt
\renewcommand{\texttt}[1]{\ifmmode\mathtt{#1}\else\VAorigtexttt{#1}\fi}

\newcommand{\VeilAudit}{\textsc{VeilAudit}}
\newcommand{\AIP}{\ensuremath{\mathcal{P}_{\mathrm{AIP}}}}
\newcommand{\AUD}{\ensuremath{\mathcal{P}_{\mathrm{AUD}}}}
\newcommand{\IRP}{\ensuremath{\mathcal{P}_{\mathrm{IRP}}}}

\newcommand{\Enc}{\ensuremath{\mathsf{Enc}}}

\newcommand{\Com}{\ensuremath{\mathsf{Com}}}
\newcommand{\ET}{\ensuremath{\mathsf{ET}}}

\newcommand{\pk}{\ensuremath{\mathsf{pk}}}
\newcommand{\sk}{\ensuremath{\mathsf{sk}}}
\newcommand{\tpk}{\ensuremath{\mathsf{tpk}}}

\newcommand{\apk}{\ensuremath{\mathsf{apk}}}
\newcommand{\tk}{\ensuremath{\mathsf{tk}}}

\newcommand{\UID}{\ensuremath{\mathsf{UID}}}
\newcommand{\Atag}{\ensuremath{\mathsf{Atag}}}

\newcommand{\CTlink}{\ensuremath{\mathsf{CT}^{\mathrm{link}}}}

\allowdisplaybreaks[1]

\date{}
\title{\Large \bf VeilAudit: Breaking the Deadlock Between Privacy and Accountability   Across Blockchains}

\author[1]{Minhao Qiao}
\author[1,2]{Hai Dong\thanks{Corresponding author: \href{mailto:hai.dong@rmit.edu.au}{hai.dong@rmit.edu.au}}}
\author[1]{Iqbal Gondal}
\affil[1]{RMIT University, Australia}
\affil[2]{Green Cryptocurrency Joint Research Laboratory (GreenCryptoLab)}
\affil[ ]{\textit{Email:} \href{mailto:S4042598@student.rmit.edu.au}{minhao.qiao@student.rmit.edu.au} \quad
\href{mailto:hai.dong@rmit.edu.au}{hai.dong@rmit.edu.au} \quad
\href{mailto:iqbal.gondal@rmit.edu.au}{iqbal.gondal@rmit.edu.au}}

\begin{document}
\maketitle

\begin{abstract}
Cross-chain interoperability in blockchain systems exposes a fundamental tension between user privacy and regulatory accountability.
Existing solutions enforce an "all-or-nothing" choice between 
full anonymity and mandatory identity 
disclosure,
limiting adoption in regulated financial settings.
We propose \textsc{VeilAudit}, a cross-chain auditing framework that 
introduces "Auditor-Only Linkability" (AOL), allowing auditors to link transaction behaviors originating from the same anonymous entity without 
learning
its
identity. 
VeilAudit achieves this via
a user-generated "Linkable Audit Tag", which 
embeds
a zero-knowledge proof to attest to its validity without exposing the user's master wallet address
and a special ciphertext 
that
only designated auditors can
test for linkage.
To balance privacy with compliance, VeilAudit further supports
a threshold-gated
identity revelation under due process. 
\textsc{VeilAudit} provides a mechanism for building reputation in pseudonymous environments, which enables applications such as cross-chain credit scoring based on verifiable behavioral history. We 
formalize
its security 
guarantees
and 
develop
a 
prototype 
spanning
multiple EVM chains. Our evaluation 
demonstrates that our framework is 
practical
for today's multi-chain environment. (Submitted to Usenix security 2026 cycle 1 in August 2025)
\end{abstract}

\section{Introduction}

The proliferation of cross-chain technologies is rapidly transforming the digital landscape, breaking down the silos between once-isolated blockchain platforms\cite{wang2023exploring,belchior2021survey}. By enabling the secure transfer of data and assets across heterogeneous networks, these interoperability protocols are unlocking novel applications in fields from finance to supply chain management\cite{qiao2025blockchain}. However, this explosion in connectivity has set two fundamental ideals of the blockchain ecosystem on a collision course: the user's demand for robust privacy and the systemic need for accountability\cite{buterin2024blockchain}. As users rightfully adopt powerful privacy-enhancing technologies, the transparency required for auditing and regulatory compliance is critically undermined, creating a foundational challenge for the future of decentralized systems.

This tension is not merely a philosophical debate but a deep-seated technical conflict. Modern privacy-preserving mechanisms, such as zero-knowledge proofs (ZKPs)\cite{liang2025sok,chaliasos2024sok}, transaction mixing\cite{heilman2017tumblebit,ni2024cmixing}, and ring signatures\cite{lai2019omniring}, are expressly designed to obfuscate transaction details and break the chain of provenance. While highly effective at protecting user anonymity, these techniques render traditional auditing paradigms—which rely on ledger transparency—ineffective\cite{narula2018zkledger}c. The problem is dangerously amplified in the cross-chain context. A transaction may traverse multiple autonomous domains, each with its own identity and privacy standards, creating a fragmented and obscured trail that is nearly impossible for an external observer to reconstruct. This makes crucial tasks like transaction link analysis, behavioral tracing, and responsibility attribution exceptionally difficult.

The consequences of this unresolved conflict are severe, creating a major barrier to mainstream adoption. In finance, the lack of auditable accountability undermines the credibility of cross-chain assets for institutional investors, impedes the growth of innovative DeFi applications like under-collateralized lending, and inadvertently creates safe havens for illicit activities such as money laundering\cite{bartoletti2021sok,miedema2023mixed}. Without a framework that can reconcile robust privacy guarantees with the practical needs for audit and accountability, the broader vision of a trustworthy and compliant decentralized economy remains unrealized.

To bridge this critical gap, a paradigm shift is needed in how we approach on-chain accountability. A viable solution must be built from the ground up to respect user privacy by default, while providing the necessary tools for authorized oversight. This paper aims to design such a system by addressing the core research problems at the intersection of cryptography, distributed systems, and regulatory compliance.

\noindent
\textbf{Research Problems.} The fundamental question is: how can one design an auditing and accountability framework that preserves the strengths of secure, private, and unlinkable cross-chain transactions? This challenge manifests in four core aspects:

\noindent\textbf{(1) Minimal Disclosure Auditing.} How can we reconcile privacy protection with auditable transparency? The system must allow auditors to validate transaction legality and compliance without revealing any sensitive content or user identity. In particular, we ask: how can one support both \textit{identity unlinkability} and \textit{auditing linkability}, such that multiple transactions on different chains from the same entity can be linked by the auditor (under certain conditions), without revealing the entity itself?

\noindent\textbf{(2) Auditing Completeness and Efficiency.} In practice, auditors operate offline and across different time windows. They must retrospectively analyze user behaviors, often across multiple blockchains and time epochs. Therefore, audit data must be persistently available, queryable, and verifiable, without requiring interactive communication with the transaction originator. Moreover, the auditing process must scale efficiently across large datasets and heterogeneous sources (from different chains).

\noindent\textbf{(3) Controlled De-anonymization.} Regulatory authorities often require more than just proof of transaction validity—they demand mechanisms for identifying transactors under legal authorization. This requires the design of a controlled de-anonymization mechanism, where user identities can only be revealed when certain thresholds or legal conditions are met. Such mechanisms must prevent unauthorized identity exposure and ensure full accountability and audit traceability.

\noindent\textbf{(4) Low Overhead and Performance Awareness.} Cross-chain systems are inherently resource-constrained: blockchain throughput is limited, and inter-chain communication adds latency. While recursive zero-knowledge proofs provide strong privacy guarantees, they introduce prohibitive costs. Thus, we ask: how can one design privacy-preserving audit mechanisms with minimal computational and bandwidth overhead? 

A related line of work, zkCross\cite{guo2024zkcross}, introduces protocols that primarily targets privacy-preserving cross-chain identity authentication in audit. Its design focuses on enabling users to prove their state on a source chain when interacting with a target chain, while maintaining anonymity. However, zkCross does not address the auditability and accountability of transaction behaviors. In contrast, our framework is centered on transaction-level verifiability and responsibility attribution, enabling the recording, reconstruction, and aggregation of user behaviors across heterogeneous chains. This allows for compliance enforcement and risk control at the behavioral level across domains. While zkCross emphasizes anonymity and state privacy, it lacks the capability to determine whether multiple transactions originate from the same user—an essential feature for behavior clustering or entity recognition in compliance scenarios. Moreover, our framework supports threshold-controlled de-anonymization, whereby the real identity associated with a transaction can be revealed only under authorized conditions defined by the auditing chain. This mechanism satisfies the demands of financial regulation and accountability without compromising user privacy under normal operations.

\subsection{VeilAudit Approach}

VeilAudit resolves the privacy–accountability conflict
by shifting the focus from hiding transactions to enabling the linkage of anonymous actions without revealing the master wallet addresses. Our approach avoids any form of on-chain identity registration. Instead, it introduces a novel, post-transaction auditing mechanism centered around a unique cryptographic artifact: the Linkable Audit Tag.

The core of our approach lies in a two-step process for constructing this Tag after a transaction is finalized:

First, to execute a transaction, the user generates a temporary, one-time virtual address. The user then employs a zero-knowledge proof (zk-SNARK) in a particularly novel way: not to hide the transaction amount, but to generate a cryptographic proof of ownership. This proof attests that the user controls the master wallet address that derived this temporary address, all without computationally linking the two or revealing the master wallet address itself. This step assures the auditor that the Tag is authentically generated by a legitimate, key-holding entity.

Second, embedded within each Tag is a specialized ciphertext called equality-test ciphertext ($\CTlink$). This component is created using Public-key Encryption with Equality Test (PKE-ET). It functions as a unique cryptographic lockbox: an authorized auditor holds the exclusive key to test if any two lockboxes originate from the same anonymous user. Crucially, they can verify this link but can never open the box to see the user's actual master wallet address.

This two-part design allows auditors to collect Tags on a dedicated audit chain and construct a behavioral graph of an anonymous entity's cross-chain activities. They can finally break the "all-or-nothing" privacy deadlock—enabling effective compliance through linkability auditing, while rigorously protecting the anonymity of all users.

\subsection{VeilAudit Features}

The VeilAudit approach introduces several features designed to reconcile the conflicting requirements of privacy and compliance in the multi-chain world.

First, it enables Auditor-Only Linkability (AOL). This is the core benefit. It creates a new paradigm where auditors can verify that Tag A on Ethereum and Tag B on Solana were created by the same anonymous entity, allowing them to detect sophisticated illicit patterns like cross-chain wash trading or money laundering. For all other parties, including other users and the public, these tags remain completely unlinked and anonymous, thus protecting herd privacy.

Second, it offers a Resource-Efficient, Non-Interactive Auditing model. Audit Tags are generated post-hoc, meaning they are only created after a transaction has been successfully confirmed on-chain. This is highly efficient, as no on-chain resources are wasted on failed or pending transactions. Furthermore, once a Tag is submitted, the process is non-interactive; auditors can perform their analysis without any further input from the user, enhancing both efficiency and privacy.

Third, VeilAudit is designed with High Adaptability for the real world. Its architecture is decoupled, using a relay bridge module to separate the audit layer from the execution layer. This modular design makes it compatible with all mainstream cross-chain technologies and allows for easy integration into existing infrastructure without requiring fundamental changes to the underlying blockchains. (Our proof-of-concept validates this by integrating with EVM chains via the Axelar bridge).

Finally, it provides a Threshold-Controlled Accountability Mechanism as a crucial safeguard. For situations where a pattern of activity is deemed highly suspicious and legally actionable, VeilAudit includes a controlled de-anonymization mechanism. This is not a backdoor. Identity revelation can only be triggered under strict, pre-defined conditions through an on-chain collaborative authorization involving multiple, independent parties (e.g., a court, a regulatory body, and a project DAO). This entire process is transparently logged, ensuring accountability is possible without enabling unauthorized access or misuse of power.

\subsection{Application Scenarios: Enabling a Cross-Chain Reputation and Credit}

A key application of VeilAudit is enabling portable behavioral credentials across blockchains. For instance, a user’s history of repaying loans on Ethereum can be cryptographically linked and verified on Solana without revealing their identity. This addresses the “reputation black hole” in DeFi, reducing the need for extreme over-collateralization or off-chain identity disclosure. VeilAudit's Linkable Audit Tags are a practical implementation of such a credential.

This capability is critical for solving a major challenge in today's cross-chain DeFi landscape: the "Reputation Black Hole." Currently, a user's valuable on-chain history is trapped within a single blockchain. A user who has built up a stellar reputation by consistently repaying loans on Ethereum is treated as a complete unknown—and therefore untrustworthy—when they bridge assets to a lending protocol on Solana.

To mitigate this risk, platforms are forced into two undesirable extremes:

\noindent(1) Extreme Over-collateralization: They demand excessively high collateral from all anonymous users, leading to massive capital inefficiency across the ecosystem.

\noindent(2) Permissioned "Walled Gardens": They abandon the ethos of decentralization and require users to complete off-chain KYC, linking their real-world identity to their wallets, which completely sacrifices user privacy.

VeilAudit resolves this dilemma. A user can leverage VeilAudit to generate a Portable Behavioral Credential from their Ethereum transaction history. This credential, composed of linked, anonymous audit tags, cryptographically proves a consistent pattern of good behavior (e.g., "this entity has successfully closed 50 loan positions"). They can then present this credential to the Solana lending protocol. The protocol can verify the integrity of this behavioral history and, based on this newly established trust, offer the user more favorable terms, such as lower collateral requirements.

In essence, VeilAudit enables the creation of a trustworthy pseudonymous economy. Reputation becomes a portable asset, fostering greater capital efficiency and more sophisticated risk management in DeFi. This same principle extends to other critical areas, such as allowing regulators to verify compliance paths in cross-border finance without pre-emptive identity disclosure, or enabling judicial bodies to attribute responsibility in digital forensics under strict, authorized conditions.

\subsection{Contributions}
In summary, our contributions are as follows:

\noindent(1) We propose VeilAudit, the first general-purpose framework for cross-chain auditing that resolves the structural tension between user privacy and regulatory accountability. Our system introduces the concept of "Auditor-Only Linkability" (AOL), enabling behavioral analysis and risk management in anonymous environments without revealing master wallet addresses.

\noindent(2) We formalize the cryptographic core of VeilAudit by introducing Linkable Audit Tags (LATs). We present a novel construction that combines zero-knowledge proofs of ownership with Public-key Encryption with Equality Test (PKE-ET) to create verifiable, portable, and pseudonymous behavioral credentials. We also provide a security analysis for our construction.

\noindent(3) We present a practical, proof-of-concept implementation of VeilAudit, demonstrating its feasibility in a real-world multi-chain environment (linking two EVM chains via a Cosmos-based audit chain). Our decoupled and modular architecture ensures high adaptability and resource efficiency, making VeilAudit a practical solution for the diverse cross-chain ecosystem.

\noindent
\textbf{Structure of the Paper.} The remainder of this paper is organized as follows. 
Section~\ref{sec:Preliminaries} presents the necessary preliminaries underlying our work. 
In Section~\ref{sec:Framework}, we introduce the general auditing architecture on which VeilAudit is built. 
Section~\ref{sec:Model} formalizes the complete VeilAudit system, detailing its architecture, workflow, threat model, and the core cryptographic protocols it is built upon.
Section~\ref{sec:security} provides a comprehensive security analysis of the entire system and its protocols. 
Performance evaluation results are reported in Section~\ref{sec:Evaluation}. 
We discuss related work in Section~\ref{sec:related}, and finally conclude the paper and outline future research directions in Section~\ref{sec:conclusion}.

\section{Framework}
\label{sec:Framework}

\subsection{Architectural Principles and Functional Decomposition}
\label{subsec:architecture}

To address the dual goals of \textit{privacy-preserving cross-chain behavior tracing} and \textit{regulated accountability}, we propose a general-purpose framework for auditable systems that can operate over heterogeneous blockchains and interoperability protocols. This architecture is not tied to any specific implementation or protocol instance; rather, it is intended to serve as a \textbf{reusable design pattern} that other systems can adopt and extend. The framework organizes the system into three distinct but interoperable layers:

\noindent \textbullet\phantom{ } \textbf{Transaction Layer:} This layer encompasses all application-specific logic and user-initiated transactions on existing blockchains. It assumes no native support for identity or audit logic. Instead, users interact using anonymous addresses derived from local keys, ensuring unlinkability by default. This layer is agnostic to how auditability is achieved.

\noindent \textbullet\phantom{ }
\textbf{Audit Layer:} Sitting orthogonally to the base chains, this layer ingests verifiable metadata from finalized transactions (e.g., control proofs, provenance signals, identity commitments). It verifies and stores audit tags that support \textit{equality testing} and \textit{behavioral linkage}, without disclosing identities. The layer can be instantiated as a standalone blockchain, off-chain oracle network, or privacy-preserving storage system.

\noindent \textbullet\phantom{ }
\textbf{Accountability Layer:} This top layer governs \textit{controlled identity recovery}, typically via \textit{threshold-decryptable capsules} submitted alongside audit metadata. It also enforces due process policies, such as legal threshold conditions and immutable proof trails. This layer is activated only when accountability is required.

This functional separation promotes \textbf{modularity, upgradability, and minimal disclosure}, allowing each layer to evolve independently. Within this model, key actors have clearly defined roles. \textbf{Users} operate on the Transaction Layer. \textbf{Auditors} are the primary actors on the Audit Layer, empowered by its equality-testing capabilities. An \textbf{Accountability Authority} (e.g., an Unmasking Committee) is the designated entity for the Accountability Layer, whose power is constrained by a foundational cryptographic setup, such as a \textbf{threshold public key ($\tpk$)}.

\subsection{Technical Challenges}
\label{sec:core-challenges}

To achieve privacy-preserving cross-chain auditing with regulated accountability, the VeilAudit framework is designed to solve three fundamental and interrelated technical challenges. This section details each challenge and presents our cryptographic solutions.

\subsubsection*{Challenge 1: Constructing Verifiable yet Privacy-Preserving Audit Tags}
\label{subsec:challenge-tags}

The foundational component of our system is the \textit{audit tag}, a piece of metadata that must be verifiable, unforgeable, and privacy-preserving. The core challenge is to design a tag that satisfies three distinct requirements: proving its legitimacy without revealing the user's identity; ensuring it is bound to a real transaction to prevent forgery; and enabling an auditor to link a user's activities while keeping the identity unlinkable and private even from the auditor.

Our solution is a composite tag that addresses each requirement with a specific cryptographic primitive:

\noindent\textbf{(1) Legitimacy via Single-Use Anonymous Identities.} To prove a tag’s origin without exposing public keys, we use a zk-SNARK to generate a \textbf{single-use anonymous identity} for each transaction. The proof attests that this new identity was derived from a valid, user-controlled signing key, but the identity itself is computationally unlinkable to the user's public address or any other generated identity. This provides a cryptographically authentic yet ephemeral persona for each audit event, confirming the tag's legitimacy while preserving long-term anonymity.

\noindent\textbf{(2) Integrity via Transaction Binding.} To prevent forgery and replay attacks, each audit tag is structurally and immutably bound to a finalized on-chain transaction. This is achieved by embedding a cryptographic commitment (e.g., a hash) to the unique transaction record within the tag. The audit chain only accepts tags corresponding to a real, irreversible transaction, thus eliminating the possibility of synthetic tag injection.

\noindent\textbf{(3) Private Linkability via Encryption with Equality Test (PKE-ET).} To solve the paradox of enabling auditor-only linkage while maintaining anonymity, we employ Public-key Encryption with Equality Test. A derivative of the user's hidden master identity is encrypted under the auditor's public key ($\apk$) to generate a ciphertext, $\CTlink$. While appearing as random, unlinkable data to the public, the auditor can use their exclusive trapdoor key ($\tk$) to test if any two $\CTlink$ values correspond to the same underlying identity. This allows the auditor to link behavior across chains while the user's identity remains cryptographically unexposed.

\subsubsection*{Challenge 2: Ensuring Audit Integrity in a Trustless Cross-Chain Environment}
\label{subsec:challenge-integrity}

In a decentralized and asynchronous environment, we cannot assume users will cooperate with audits. The challenge is to design a system that ensures the audit trail is complete, objective, and resistant to manipulation (like replay attacks) without requiring any user interaction post-transaction.

Our solution is a non-interactive process that ensures data integrity and low overhead:

\noindent\textbf{(1) Non-Interactive Auditing for Completeness.} The client-side component generates the audit tag and the accompanying zero-knowledge proof \textbf{locally and automatically} during the transaction submission process. This enables auditors to verify compliance offline and asynchronously, ensuring a complete audit trail without relying on user participation.

\noindent\textbf{(2) Uniqueness Verification for Replay Resistance.} The binding of each tag to a unique transaction hash and timestamp is crucial for mitigating replay attacks. The audit chain enforces a strict "one tag per transaction" rule by performing deduplication checks, preventing an attacker from polluting the audit data by repeatedly broadcasting information related to a single transaction.

\noindent\textbf{(3) Low-Overhead Primitives for Practicality.} To ensure the system is deployable in resource-constrained blockchain environments, the design incorporates lightweight and modular cryptographic primitives. For example, we opt for an efficient \textbf{on-chain signature challenge} to prove control of funds, avoiding the high gas costs associated with verifying ECDSA signatures inside a ZK circuit.

\subsubsection*{Challenge 3: Balancing Anonymity with Controlled De-anonymization}
\label{subsec:challenge-deanonymization}

A purely anonymous system is incompatible with regulatory requirements. The challenge is to preserve user anonymity by default, yet enable controlled, auditable identity revelation under legally authorized conditions, ensuring that this powerful capability cannot be abused or create new risks for the user.

Our solution is a cryptographically-enforced, governance-driven de-anonymization mechanism:

\noindent\textbf{(1) Identity Escrow via Threshold Encryption.} Each user's master public key ($pk_{\mathsf{master}}$) is encrypted under a \textbf{threshold public key} ($\tpk$) held by a decentralized committee of authorities. No single party holds the complete key to decrypt this identity.

\noindent\textbf{(2) Governance-Gated Identity Recovery.} De-anonymization can only be triggered by a formal, on-chain governance process, requiring the submission of valid audit evidence and a successful vote by a quorum ($t$-of-$n$) of the authority committee. Only when a threshold number of members provide their decryption shares can the original identity be reconstructed.
    
\noindent\textbf{(3) Post-Revelation Security.} The accountability mechanism is designed to prevent user impersonation. The revealed identity ($pk_{\mathsf{master}}$) serves as a unique identifier for legal purposes but is cryptographically distinct from the operational signing keys used for daily transactions. Therefore, even an adversary who obtains the revealed public key through the regulated process cannot use it to forge signatures or control the user's funds. This ensures that the de-anonymization process does not introduce new attack vectors against the user.

\section{Model}
\label{sec:Model}
\subsection{Network Model}

VeilAudit adopts a two-layer blockchain architecture comprising a set of \textbf{transaction chains (Layer-1)} and a dedicated \textbf{audit chain (Layer-2)}. The two layers are interconnected via a cross-chain communication protocol, Axelar, which facilitates message relay and state synchronization in a secure and decentralized manner.

\textbf{Layer-1:} The transaction layer consists of multiple independent, Turing-complete public blockchains. All chains in this layer are required to support smart contract execution and zero-knowledge proof primitives, enabling expressive interactions and verifiable computation. In the current implementation, VeilAudit utilizes EVM-compatible blockchains (e.g., private Ethereum instances) to ensure compatibility with existing infrastructure and tooling. Users perform transactions using their native wallet accounts, and all transactional data remains on-chain and accessible for subsequent referencing.

\textbf{Layer-2:} The audit layer is instantiated as an independent blockchain built using the Cosmos SDK. It is responsible for receiving and recording structured audit data transmitted from the transaction layer. The audit chain is designed to be lightweight and modular, with support for permission control and multi-party threshold mechanisms. It remains logically decoupled from the transaction chains, and does not participate in transaction execution or consensus formation within Layer-1.

\textbf{Cross-Chain Communication.} To bridge the two layers, VeilAudit integrates the Axelar cross-chain protocol, which provides light-client-based message validation and cryptographic signature forwarding. Bridge relayers operating in the Axelar network observe specific events on the transaction chains and encapsulate them as standardized cross-chain messages. These messages are then authenticated and submitted to the audit chain. Bridge relayers function purely as transport intermediaries, with no ability to modify, forge, or influence the content or status of the underlying transactions.

\subsection{Threat Model}
\label{sec:threatmodel}

\paragraph{Participants and Trust Assumptions.}
The VeilAudit system consists of the following entities:
\[
\mathcal{E} = \{\mathcal{U}, \mathcal{R}, \mathcal{A}, \mathcal{C}\}
\]

\noindent \textbullet\phantom{ } \textbf{Users} $\mathcal{U}$: All public participants capable of initiating transactions on Layer-1 (the transaction chains). Since the audit chain may be permissionless, $\mathcal{U}$ includes any on-chain actor without prior registration or approval. After a transaction is completed, users are assumed to abstain from any follow-up interaction or verification and may actively evade audit. They are modeled as \emph{malicious} adversaries.

\noindent \textbullet\phantom{ }
\textbf{Relayers} $\mathcal{R}$: Bridge components under the Axelar protocol responsible for monitoring Layer-1 events and forwarding verified messages to Layer-2. Relayers are assumed to be \emph{trusted under normal network conditions}, meaning they do not modify, drop, or forge messages, though temporary delays and interruptions are tolerated.

\noindent \textbullet\phantom{ }
\textbf{Auditors} $\mathcal{A}$: Entities operating on Layer-2 (the audit chain) to retrieve and analyze tag data from Layer-1. Auditors are modeled as \emph{honest-but-curious}: they follow the protocol but may attempt to infer user identities or behavioral linkages by inspecting tag content and querying public Layer-1 data.

\noindent \textbullet\phantom{ }
\textbf{Authority Committee} $\mathcal{C}$: A set of compliance-approved organizations holding threshold decryption keys. This set is modeled under a \emph{Byzantine fault-tolerant majority assumption}: identity recovery from a tag is only possible if $t \leq |\mathcal{C}|$ parties jointly authorize the operation.

\paragraph{Adversary Model.}
We define the adversary set as:
\[
\mathcal{A}dv = \{\mathcal{A}dv_{\mathcal{U}},~\mathcal{A}dv_{\mathcal{A}}\}
\]
where $\mathcal{A}dv_{\mathcal{U}}$ represents malicious users, and $\mathcal{A}dv_{\mathcal{A}}$ denotes curious auditors.

\noindent\phantom{ }
\textbf{Attack I: Tag Forgery.}
\label{attack:forgery}
A malicious user attempts to fabricate a valid-looking tag \(\tau'\) that is not derived from any legitimate transaction. The goal is to mislead the audit chain into recording falsified data.

\[
\exists~\tau' \notin \mathcal{T}_{\text{valid}} \text{ such that } \texttt{Verify}(\tau') = \texttt{True}
\]

\noindent\phantom{ }
\textbf{Attack II: Tag Replay.}
\label{attack:replay}
An adversary captures a valid tag \(\tau\) and submits it multiple times to the audit chain to pollute records or confuse correlation algorithms.
\[
\tau \in \mathcal{T}_{\text{valid}}, \quad \texttt{Count}(\tau) \gg 1
\]
\noindent\phantom{ }
\textbf{Attack III: Auditor-driven Tag-to-Identity Inference.}
\label{attack:linkage-inference}
An honest-but-curious auditor uses public data from Layer-1 to establish a link between a given tag \(\tau\) and an on-chain address \(\texttt{addr}_{\text{L1}}\).
\[
\texttt{Link}(\tau) \rightarrow \texttt{addr}_{\text{L1}} \quad \text{with } \Pr > \varepsilon
\]
where \(\texttt{Link}(\cdot)\) uses metadata such as timestamp, interaction pattern, or contract behavior to perform inference.

\noindent\phantom{ }
\textbf{Attack IV: Cross-Tag Correlation and Clustering.}
\label{attack:cross-tag-cluster}
An auditor aggregates multiple tags and attempts to group them under a common pseudonymous identity, undermining unlinkability.
\[
\exists~\mathcal{T}' \subset \mathcal{T} \text{ such that } \texttt{Cluster}(\mathcal{T}') \rightarrow \mathcal{U}_i
\]
\noindent\phantom{ }
\textbf{Attack V: Unauthorized De-anonymization.}
\label{attack:unauth-reveal}
An adversary attempts to bypass the threshold decryption policy and extract the original user identity bound to a tag without proper authorization.
\[
\texttt{Reveal}(\texttt{addr}_{\text{origin}}~|~\tau) \quad \text{without } t\text{-out-of-}|\mathcal{C}|
\]
\noindent\phantom{ }
\textbf{Attack VI: Post-Disclosure Identity Impersonation.}
\label{attack:post-disclosure}

After a legitimate regulatory process reveals a user's master public key \(\texttt{pk}_{\text{master}}\), an adversary attempts to impersonate that user by generating valid audit tags or control proofs without knowing the corresponding master private key \(\texttt{sk}_{\text{master}}\): 
\[
\mathsf{Forge}(\texttt{pk}_{\text{master}}) \Rightarrow \texttt{Atag}^* \ \text{or} \ \pi_{\text{ctrl}}^* \quad \text{s.t.} \quad \texttt{Verify}(\cdot) = \texttt{True}
\]
\subsection{VeilAudit Overview}

\begin{figure*}[t]  
 \vspace{-20pt} 
  \centering
  \includegraphics[width=\textwidth]{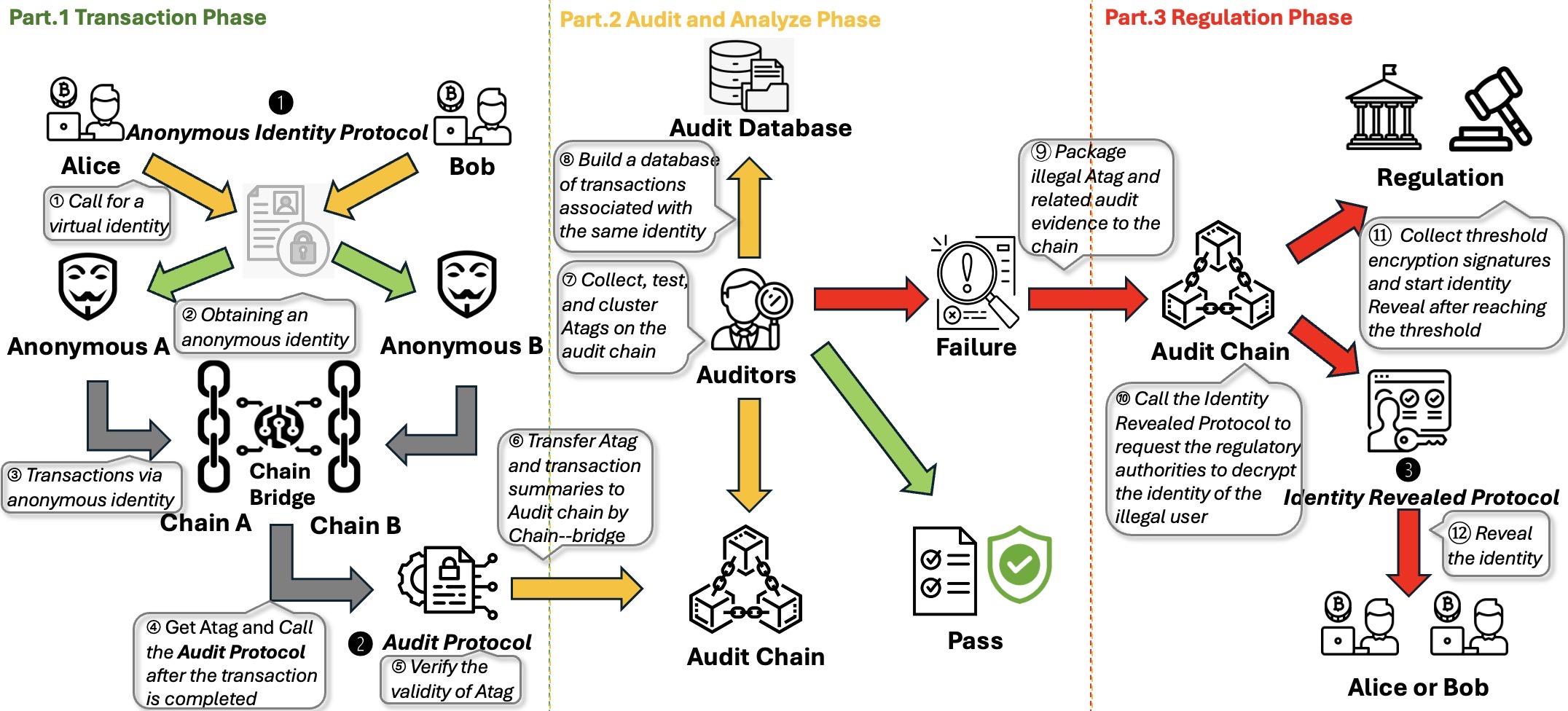} 
  \caption{System Overview of VeilAudit.}
  \label{fig:overview}
  \vspace{-15pt} 
\end{figure*} 

VeilAudit aims to reconcile privacy preservation with accountable cross-chain auditing by adopting a layered system architecture and a three-phase operational workflow. An overview of the full architecture and workflow is illustrated in Figure~\ref{fig:overview}. As a concrete instantiation of the abstract auditing framework proposed in Section~\ref{sec:Framework}, VeilAudit demonstrates how key functional roles and protocol boundaries can be realized under real-world cryptographic and deployment constraints. Within this model, three core protocols—namely, the Anonymous Identity Protocol~(\(\mathcal{P}_{\text{AIP}}\)), the Audit Protocol~(\(\mathcal{P}_{\text{AP}}\)), and the Identity Revealed Protocol~(\(\mathcal{P}_{\text{IRP}}\))—are invoked at different stages to accomplish unlinkable identity generation, audit label construction, and threshold-based identity disclosure under regulatory supervision. This section provides a comprehensive overview of the system’s full-process execution model, while protocol-level details are deferred to Section~\ref{sec:protocols}.

\paragraph{Transaction Phase.}

In the transaction phase of \textit{VeilAudit}, two users, Alice and Bob (denoted as \( \mathcal{U}_A \) and \( \mathcal{U}_B \), respectively), initiate a privacy-preserving cross-chain transaction on their respective Layer-1 public blockchains. \Circled{1} To protect user anonymity, the system first invokes the \textit{Anonymous Identity Protocol} \( \mathcal{P}_{\text{AIP}} \). \Circled{2} Under this protocol, each user generates a new unlinkable anonymous address, \( \texttt{addr}_A^{\text{anon}} \) and \( \texttt{addr}_B^{\text{anon}} \), which is solely controlled by its owner and unlinkable to prior identities.

To ensure the validity of these anonymous addresses, each user generates a zero-knowledge proof of control \( \pi_{\text{ctrl}}^A \) and \( \pi_{\text{ctrl}}^B \), demonstrating ownership of the corresponding anonymous address without disclosing their original public keys or identities.

For asset preparation, users deposit the intended transaction value \( \texttt{val}_A \) and \( \texttt{val}_B \) into neutral escrow contracts \( \texttt{Esc}_A \) and \( \texttt{Esc}_B \) deployed on their respective chains. These escrow contracts are system-level reusable containers not bound to any specific user. Once \( \mathcal{P}_{\text{AIP}} \) verifies that both users have sufficient balances and valid control proofs, it instructs the escrows to transfer the designated amount into the anonymous addresses \( \texttt{addr}_A^{\text{anon}} \) and \( \texttt{addr}_B^{\text{anon}} \). \Circled{3} The subsequent cross-chain transfer is then executed between \( \texttt{addr}_A^{\text{anon}} \) and \( \texttt{addr}_B^{\text{anon}} \), coordinated via a cross-chain messaging bridge set \( \mathcal{R} \) such as Axelar.

Concurrently, the protocol captures the users’ original public keys \( \texttt{pk}_A \) and \( \texttt{pk}_B \), and encrypts them via threshold encryption, resulting in ciphertexts \( \texttt{UID}_A = \texttt{Enc}_{\texttt{tpk}}(\texttt{pk}_A) \) and \( \texttt{UID}_B = \texttt{Enc}_{\texttt{tpk}}(\texttt{pk}_B) \). These ciphertexts are not broadcast on-chain but are reserved for later inclusion in audit metadata.

The actual cross-chain transaction is then executed by the anonymous addresses \( \texttt{addr}_A^{\text{anon}} \) and \( \texttt{addr}_B^{\text{anon}} \), resulting in a two-part transaction object \( t = (t_{\text{L1}A}, t_{\text{L1}B}) \).

Upon confirmation of \( t \) on-chain, the system triggers the \textit{Audit Protocol} \( \mathcal{P}_{\text{AP}} \). \Circled{4} This protocol generates a zero-knowledge execution proof \( \pi_{\text{exec}} \), which attests that the cross-chain transaction \( t \)—comprising actions from both anonymous users—has been correctly and completely executed according to protocol semantics. \( \pi_{\text{exec}} \) serves as a global post-facto proof that verifies the correctness of the entire execution trace. \(\CTlink_A, \CTlink_B\) are unlinkable identifiers derived for each user \( \mathcal{U}_A \) and \( \mathcal{U}_B \), respectively, serving as anchors for behavioral correlation.

The system then constructs an audit tag \( \texttt{Atag} \) as follows:\[
\texttt{Atag} = \mathcal{P}_{\text{AP}}(t, \pi_{\text{exec}},\pi_{\text{link}}, \CTlink_A, \CTlink_B, \texttt{UID}_A, \texttt{UID}_B)
\]

The tag \( \texttt{Atag} \) binds to a unique transaction hash \( \texttt{txid} \) and timestamp \( \texttt{ts} \), serving as the canonical audit record. \Circled{5} This tag is submitted to the audit chain (Layer-2) for future regulatory analysis under privacy-preserving conditions.

\paragraph{Audit and Analyze Phase.}

\Circled{6} The audit tag \( \texttt{Atag} \) is submitted to the audit chain (Layer-2) via the bridge \( \mathcal{R} \) (Axelar) and received by the auditor set \( \mathcal{A} \). \Circled{7} - \Circled{8} Each auditor uses their private audit key \( \texttt{sk}_{\mathcal{A}} \) to decrypt the encrypted pseudonyms \( \texttt{UID}_A \), \( \texttt{UID}_B \), enabling two levels of audit analysis:

\noindent \textbullet\phantom{ } \textbf{Single-tag analysis:} Individual tags are assessed for transaction compliance, anomaly detection, or risk profiling.

\noindent \textbullet\phantom{ } \textbf{Cross-tag behavioral correlation:} Auditors test whether multiple tags share the same encrypted pseudonym, identifying behavioral clusters without revealing user identity.

\Circled{9} If the auditors detect policy violations or suspect illicit behavior based on these analyses, they formally escalate the case to the Regulation Phase. The escalation includes submitting the implicated \( \texttt{Atag} \) entries, associated risk evidence, and compliance justification to the regulation layer through a structured on-chain request. This process complies with pre-established audit rules and acts as a trigger for initiating threshold-based identity recovery under the oversight of regulatory authorities.

\paragraph{Regulation Phase}

\Circled{10} Upon the submission of a verifiable identity-reveal request from the auditors, the final phase of VeilAudit---the \textit{Regulation Phase}---is activated. This phase is governed by a set of regulatory authorities modeled as \( \mathbb{G} = \{ \mathcal{G}_1, \mathcal{G}_2, \dots, \mathcal{G}_n \} \), where each \( \mathcal{G}_i \) denotes an independently authorized regulator participating in the threshold decryption protocol.

Upon receiving a claim request including the target audit tag \( \texttt{Atag} \), the corresponding identity-related zero-knowledge proofs, and the behavioral evidence set \( \mathcal{E}_{\text{audit}} \), each \( \mathcal{G}_i \) verifies the integrity and validity of the submission under its own policy constraints. These include evaluating the consistency of \( \pi_{\text{exec}} \), checking the correlation structure of \( \texttt{UID}_A, \texttt{UID}_B \), and ensuring the behavioral traceability warrants identity de-anonymization.

\Circled{11} If the claim passes a threshold policy defined by the system, i.e., 
$|\mathcal{G}_{\text{approve}}| \geq t, \ \mathcal{G}_{\text{approve}} \subseteq \mathbb{G}$, 
then a threshold decryption procedure is collaboratively triggered. Under this procedure, the encrypted identity ciphertexts $\texttt{CT}_A, \texttt{CT}_B$ embedded in $\texttt{Atag}$ are jointly decrypted, yielding the original public keys $\texttt{pk}_A, \texttt{pk}_B$ associated with the transaction.

\Circled{12} This de-anonymization procedure is strictly controlled by the regulatory quorum \( \mathbb{G} \) to ensure that the anonymity of users is preserved under normal circumstances and can only be bypassed when formal regulatory thresholds are satisfied. This guarantees that identity revelation occurs solely in accordance with law-enforced accountability and with resistance against unilateral abuse or insider collusion.

\subsection{Protocol}
\label{sec:protocols}

\subsubsection*{Anonymous Identity Protocol (\(\mathcal{P}_{\text{AIP}}\))}

The Anonymous Identity Protocol is responsible for initializing anonymized user identities in cross-chain transactions while preserving both unlinkability and auditability. Each user \(\mathcal{U}_i\) executes the protocol locally before engaging in any transaction, ensuring that the resulting identities are unlinkable across sessions but remain auditable in later stages.

\noindent \textbullet\phantom{ } \textbf{Anonymous Address Derivation.}  
At the beginning of each transaction session, the user \(\mathcal{U}_i\) derives a fresh anonymous key pair from its long-term master key. Specifically, the master private key \(\texttt{sk}_{\text{master}}\) is combined with a session-specific salt \(\texttt{salt}_{\text{addr}}\) through a key derivation function to produce \(\texttt{sk}_i^{\text{anon}}\). The corresponding public key yields the anonymous address \(\texttt{addr}_i^{\text{anon}}\). This process guarantees that every transaction session produces a unique address, thereby achieving unlinkability between different transactions.

\noindent \textbullet\phantom{ } \textbf{Zero-Knowledge Control Proof \(\pi_{\text{ctrl}}^i\).}  
To ensure that escrow contracts only transfer funds to legitimate anonymous addresses, the user must provide a zero-knowledge proof \(\pi_{\text{ctrl}}^i\). This proof attests that \(\mathcal{U}_i\) indeed knows the secret key associated with \(\texttt{addr}_i^{\text{anon}}\), without revealing the key itself. The escrow contract verifies \(\pi_{\text{ctrl}}^i\) before releasing funds, ensuring that only the rightful owner can initiate transactions. (See Algorithm~\ref{alg:AIP} and Circuit~\ref{circuit:ctrl} for details.)

\noindent \textbullet\phantom{ } \textbf{Execution Validity Proof \(\pi_{\text{exec}}\).}  
Once the escrow verification succeeds, the anonymous addresses execute the cross-chain transaction. To prevent replay attacks and ensure correctness, the protocol generates a proof \(\pi_{\text{exec}}\) that attests to the full execution of the cross-chain workflow under the system-defined semantics. This proof guarantees that the transaction trace---including contract calls, bridge relays, and asset transfers---is unique, authenticated, and non-replayable. (See Circuit~\ref{circuit:exec} for the construction.)

\noindent \textbullet\phantom{ } \textbf{Threshold Identity Encryption.}  
For accountability, the original public key of each user, \(pk_i\), is encrypted under the global threshold public key \(\texttt{tpk}\). The resulting ciphertext $\texttt{UID}_i = \mathsf{Enc}_{\texttt{tpk}}(pk_i)$ is unlinkable to any transaction unless a regulated decryption is later authorized. This ensures that the system can, under due process, recover identities when legally mandated.

\noindent \textbullet\phantom{ } \textbf{Commitments and Equality-Test Ciphertexts.}  
To support long-term behavioral auditing, each user commits to its master key by publishing \(\texttt{Com}_{\texttt{pk},i}\), a binding but hiding Pedersen-style commitment. In addition, the user computes an equality-test ciphertext \(\CTlink_i\) derived from the same committed identity. This ciphertext supports equality testing under a special audit key, enabling auditors to determine whether two different audit tags originate from the same hidden user. Importantly, different randomizers are used for each transaction, ensuring that the ciphertexts differ across transactions while the underlying linkage remains consistent.

\noindent \textbullet\phantom{ } \textbf{Consistency Proof \(\pi_{\text{link}}^i\).}  
To guarantee that the published values \(\texttt{UID}_i\), \(\texttt{Com}_{\texttt{pk},i}\), and \(\CTlink_i\) are mutually consistent, the user generates a zero-knowledge proof \(\pi_{\text{link}}^i\). This proof demonstrates that all three artifacts are derived from the same hidden \(pk_i\), without revealing the key itself. (See Algorithm~\ref{alg:AIP} and Circuit~\ref{circuit:link} for details.) This step ensures uniqueness of the binding: although the values differ across transactions due to randomization, they remain cryptographically tied to the same identity.

\noindent \textbullet\phantom{ } \textbf{Protocol Output.}  
The final output of \(\mathcal{P}_{\text{AIP}}\) for user \(\mathcal{U}_i\) is the tuple $\big(\texttt{addr}_i^{\text{anon}},~\pi_{\text{ctrl}}^i,~\pi_{\text{exec}},~\texttt{UID}_i,~\texttt{Com}_{\texttt{pk},i},~\CTlink_i,~\pi_{\text{link}}^i\big)$,which is later incorporated into the audit tag during the audit phase.



\subsubsection*{Audit Protocol ($\mathcal{P}_{\text{AUD}}$)}

\noindent \textbullet\phantom{ } \textbf{Inputs.}
From execution, the protocol reads (i) the finalized transaction objects on both legs, together with their chain identifiers and block metadata (transaction hashes, block hashes, heights, and confirmation depths), and (ii) a replay-resistant execution proof $\pi_{\text{exec}}$ attesting that the cross-chain message observed on the destination chain is exactly the one emitted on the source chain and that it has satisfied the bridge’s confirmation and de-duplication rules. In our deployment with Axelar, the statement proven by $\pi_{\text{exec}}$ binds the source $(\texttt{chain\_src},\texttt{txid\_src})$, the destination $(\texttt{chain\_dst},\texttt{txid\_dst})$, the bridge message identifier $\texttt{msg\_id}$, and the required confirmation depth; this prevents replays across time, forks, or chains (see Algorithm \ref{alg:AIP}).

From $\mathcal{P}_{\text{AIP}}$, the protocol receives for each user $i\in\{A,B\}$ an encrypted identity capsule $\UID_i=\Enc_{\texttt{tpk}}(\texttt{pk}_i)$, a long-term commitment $\Com_{\texttt{pk},i}$ to the user’s master public key, an equality-test ciphertext $\CTlink_i$ that enables linkability checks under a trapdoor, and a consistency proof $\pi_{\text{link}}^i$ showing that $\UID_i$, $\Com_{\texttt{pk},i}$, and the plaintext embedded in $\CTlink_i$ all originate from the same hidden identity (see Circuit \ref{circuit:link}). No secret is revealed to the audit layer by supplying these objects.

\noindent \textbullet\phantom{ } \textbf{Record construction.}
The auditor constructs one audit tag per cross-chain transfer. Conceptually, the tag contains: the source/destination chain identifiers, both transaction identifiers, a creation timestamp, the full execution proof $\pi_{\text{exec}}$, and, for each user, the tuple $(\UID_i,\Com_{\texttt{pk},i},\CTlink_i,\pi_{\text{link}}^i)$. The transaction fields make the record self-contained and temporally anchored; $\pi_{\text{exec}}$ certifies correctness and non-replayability; the per-user tuples provide privacy-preserving anchors for later clustering and (if authorized) identity recovery. No plain identities or anonymous addresses are stored inside the tag.

\noindent \textbullet\phantom{ } \textbf{Transmission and storage.}
The tag is emitted as a single cross-chain payload through the same interoperability layer (e.g., Axelar) and committed to the audit chain. On arrival, the audit contract verifies $\pi_{\text{exec}}$ and both $\pi_{\text{link}}^A,\pi_{\text{link}}^B$. Verification failure discards the tag. Successful verification appends the tag to the append-only ledger of audit records, keyed by the destination transaction identifier and the bridge message identifier. This keying, together with the statement bound by $\pi_{\text{exec}}$, enforces deduplication even under adversarial re-submission.

\noindent \textbullet\phantom{ } \textbf{Downstream use.}
Equality testing over $\CTlink_A$ and $\CTlink_B$—and across tags over time—enables auditors holding the trapdoor to perform behavioral clustering without learning identities. When a lawful unmasking is required, threshold decryption of $\UID_i$ reveals $\texttt{pk}_i$, whose hash matches the public commitment $\Com_{\texttt{pk},i}$ already stored in the tag. The combination of $\UID$, $\Com_{\texttt{pk}}$, and $\CTlink$, validated by $\pi_{\text{link}}$, guarantees that clustering and unmasking refer to the same underlying identity, while $\pi_{\text{exec}}$ guarantees that the clustered events correspond to actually executed and non-replayable cross-chain transfers.

\noindent \textbullet\phantom{ } \textbf{Output.}
The protocol outputs the on-chain audit record (the tag) on the audit chain. This record is immutable, unlinkable by default to any real-world identity, and sufficient for later accountability procedures through verification of $\pi_{\text{exec}}$ and $\pi_{\text{link}}$ and, if authorized, threshold decryption of $\UID$.


\subsubsection*{Identity Revealed Protocol ($\mathcal{P}_{\text{IRP}}$)}

\noindent \textbullet\phantom{ } \textbf{Goal and setting.}
When ex-post audit analyses identify suspicious or policy-violating behavior across a cluster of audit tags, the Identity Revealed Protocol enables \emph{regulated} de-anonymization under a $t$-of-$n$ committee threshold. The protocol discloses only the long-term master public keys of the implicated principals and only after multi-party authorization. Ordinary transactions remain anonymous and unlinkable.

\noindent \textbullet\phantom{ } \textbf{Inputs.}
A disclosure request contains a case identifier \(\texttt{caseID}\), a set of implicated audit tags \(\{\texttt{Atag}_1,\dots,\texttt{Atag}_k\}\), and the auditor’s clustering evidence indicating that these tags correspond to the same anonymous principal via equality tests on the linkable ciphertexts.  
Each \(\texttt{Atag}\) (as produced by \(\mathcal{P}_{\text{AUD}}\)) already carries: the execution proof \(\pi_{\text{exec}}\), the identity ciphertexts \(\texttt{UID}_i=\mathsf{Enc}_{\texttt{tpk}}(\texttt{pk}_i)\), the commitments \(\texttt{Com}_{\texttt{pk},i}\), the equality-test ciphertexts \(\CTlink_i\), and the consistency proofs \(\pi_{\text{link}}^i\).

\noindent \textbullet\phantom{ } \textbf{Request submission.}
The auditor submits \( (\texttt{caseID},\, \{\texttt{Atag}_\ell\}_{\ell=1}^{k},\, \text{clustering evidence}) \).
The clustering evidence states that all tags in the set are linked to the same plaintext
$m = H\!\big(\texttt{Com}_{\texttt{pk}} \,\|\, \texttt{ctx}_{\mathsf{GLOBAL}}\big)$
as witnessed by successful equality tests on the corresponding \(\CTlink\) values under the authorized trapdoor. By construction, each \(\pi_{\text{link}}^i\) (cf.\ Circuit~\ref{circuit:link}) attests that \(\texttt{UID}_i\), \(\texttt{Com}_{\texttt{pk},i}\), and \(\CTlink_i\) are derived consistently from the same hidden master identity; thus, the cluster can be interpreted as a single anonymous principal.

\noindent \textbullet\phantom{ } \textbf{Committee authorization.}
Let \(\mathbb{G}=\{\mathcal{G}_1,\dots,\mathcal{G}_n\}\) be the regulator committee with policy \(\Pi_{\text{reg}}\) and threshold \(t\).
Each member independently reviews the request against \(\Pi_{\text{reg}}\) (e.g., statutory triggers, severity, proportionality) and issues an approval or denial.
If the number of approvals reaches \(t\), the system records an authorization artifact containing the approving set and a timestamp and proceeds to threshold decryption.

\noindent \textbullet\phantom{ } \textbf{Threshold decryption.}
For each implicated party \(i\), committee members produce partial decryption shares \(\mathsf{Dec}_{\mathcal{G}_j}(\texttt{UID}_i)\).
Once at least \(t\) valid shares are collected, the system computes $\texttt{pk}_i^{\text{master}} \;\gets\; \mathsf{Combine}\big(\{\text{shares}_i\}\big),$
revealing the master public key associated with the clustered tags.
If multiple tags in the cluster correspond to the same \(\texttt{Com}_{\texttt{pk}}\), a single decryption suffices; the remaining tags are linked to the result through \(\texttt{Com}_{\texttt{pk}}\).

\noindent \textbullet\phantom{ } \textbf{Result materialization and accountabilit-y.}
The audit chain records the tuple
$\left(\texttt{caseID},\; \texttt{Atag}_\ell\,\; \text{authorization~artifact},\;
\texttt{pk}_i^{\text{master}}\right),$
and emits an \(\textsf{IdentityRevealed}\) event.
Regulators then map \(\texttt{pk}^{\text{master}}\) to off-chain KYC records and pursue actions mandated by law and policy.

\noindent \textbullet\phantom{ } \textbf{Relation to prior artifacts.}
The proof \(\pi_{\text{exec}}\) (cf.\ Circuit~\ref{circuit:exec})—already included in each \(\texttt{Atag}\)—attests that the cross-chain execution satisfied protocol semantics and anti-replay constraints.
The proof \(\pi_{\text{link}}^i\) (cf.\ Circuit~\ref{circuit:link}) guarantees the internal consistency among \(\texttt{UID}_i\), \(\texttt{Com}_{\texttt{pk},i}\), and \(\CTlink_i\).
The IRP itself introduces no new zero-knowledge verification; it only performs threshold-authorized decryption and on-chain materialization of the outcome.

\section{Security Analysis}
\label{sec:security}

Our security analysis, conducted within the Random Oracle Model (ROM), demonstrates that VeliAudit is secure against any probabilistic polynomial-time (PPT) adversary  $\mathcal{A}$. By formalizing security games and providing proof sketches for each property, we address the threat scenarios defined in Section~\ref{sec:threatmodel} (Attacks I--VI), along with an additional identity-impersonation scenario following public key disclosure. The analysis confirms that under these assumptions, VeliAudit successfully provides minimal-disclosure auditing, auditor-only linkability, unforgeable and replay-resistant audit tags, and threshold-gated identity revelation, with any adversary achieving only a negligible advantage.

\subsection{Assumptions}
We rely on the following cryptographic assumptions:\\
\textbullet\phantom{ } \textbf{zk-SNARKs:} The zk-SNARK scheme $\Pi_{\mathsf{SN}}$ satisfies completeness, soundness, and zero-knowledge.

\noindent\textbullet\phantom{ } \textbf{Hash Function:} $H$ is collision-resistant and modeled as a random oracle.

\noindent \textbullet\phantom{ }  \textbf{Commitments:} Pedersen commitments are computationally binding and perfectly hiding.

\noindent \textbullet\phantom{ } \textbf{Threshold Encryption:} Threshold ElGamal is IND-CPA secure in the $t$-out-of-$n$ setting.

\noindent \textbullet\phantom{ } \textbf{PKE-ET:} Public-key encryption with equality test is semantically secure; randomized encryption prevents public equality testing without the equality-test key.

\noindent \textbullet\phantom{ } \textbf{Threshold Signatures:} BLS threshold signatures are existentially unforgeable under chosen-message attack (EUF-CMA).

\subsection{Security Properties and Games}

\paragraph{Property 1: Minimal Disclosure Auditing (MDA).}
Given only public chain data and the set of audit tags $\{\mathsf{Atag}\}$, no PPT adversary
can compute any non-trivial function of $pk_{\mathrm{master}}$ with non-negligible probability
(see Attack~\ref{attack:linkage-inference} for the corresponding adversarial goal; full proof in Appendix~\ref{P1}):
\[
\mathsf{Adv}^{\mathrm{MDA}}_{\mathcal{A}}(\lambda) \le \operatorname{negl}(\lambda).
\]

\noindent
\textbf{Game $\mathsf{Game}^{\mathsf{MDA}}$:}
\begin{enumerate}[leftmargin=*]\setlength\itemsep{0em} 
    \item \emph{Setup:} Challenger $\mathcal{C}$ generates $(pk_{\mathsf{master}}, sk_{\mathsf{master}})$ and corresponding $\mathsf{Atag}$ values for multiple transactions.
    \item \emph{Challenge:} $\mathcal{C}$ sends all public data to $\mathcal{A}$.
    \item \emph{Guess:} $\mathcal{A}$ outputs $f(pk_{\mathsf{master}})$ for some non-trivial $f$.
\end{enumerate}
The advantage is $\Pr[f(pk_{\mathsf{master}})$ correct$] - \frac{1}{|\mathcal{R}|}$, where $\mathcal{R}$ is the range of $f$.

\noindent
\emph{Proof Sketch:} Perfect hiding of commitments and IND-CPA security of both threshold encryption and PKE-ET ensure that all values related to $pk_{\mathsf{master}}$ are computationally indistinguishable from random.

\paragraph{Property 2: Auditor-Only Linkability (AOL).}
\label{prop:aol}
Without the equality-test key $\tk$, no PPT adversary can determine whether two tags $\mathsf{Atag}_i, \mathsf{Atag}_j$ originate from the same entity with advantage better than random guessing (see Attack~\ref{attack:cross-tag-cluster} for the corresponding adversarial goal; full proof in Appendix~\ref{P2}):
\[
\mathsf{Adv}^{\mathsf{AOL}}_{\mathcal{A}}(\lambda) \leq \mathsf{negl}(\lambda).
\]

\noindent
\textbf{Game $\mathsf{Game}^{\mathsf{AOL}}$:}
\begin{enumerate}[leftmargin=*]\setlength\itemsep{0em} 
    \item \emph{Setup:} $\mathcal{C}$ generates keys and issues tags for multiple users across transactions.
    \item \emph{Challenge:} $\mathcal{C}$ picks either two tags from the same user ($b=1$) or from different users ($b=0$) and gives them to $\mathcal{A}$.
    \item \emph{Guess:} $\mathcal{A}$ outputs $b'$.
\end{enumerate}
The advantage is $|\Pr[b'=b] - 1/2|$.

\noindent
\emph{Proof Sketch:} Since $\mathsf{CT}^{\mathsf{link}}$ encrypts $m = H(\Com_{\mathsf{pk}} \parallel \mathsf{ctx}_{\mathsf{GLOBAL}})$ under randomized PKE-ET, without $\tk$ all ciphertexts are computationally indistinguishable.

\paragraph{Property 3: Audit Tag Unforgeability (ATUF).}
\label{prop:atuf}
No PPT adversary can create a valid new $\mathsf{Atag}$ without a valid Axelar threshold signature and a satisfying proof $\pi_{\mathsf{link}}$(see Attack~\ref{attack:forgery} and Attack~\ref{attack:unauth-reveal} for the corresponding adversarial goal; full proof in Appendix~\ref{P3}):
\[
\mathsf{Adv}^{\mathsf{ATUF}}_{\mathcal{A}}(\lambda) \leq \mathsf{negl}(\lambda).
\]

\noindent
\textbf{Game $\mathsf{Game}^{\mathsf{ATUF}}$:}
\begin{enumerate}[leftmargin=*]\setlength\itemsep{0em} 
    \item \emph{Setup:} $\mathcal{C}$ generates public parameters and gives them to $\mathcal{A}$.
    \item \emph{Challenge:} $\mathcal{A}$ outputs a candidate $\mathsf{Atag}^*$.
    \item \emph{Verify:} $\mathsf{Atag}^*$ is accepted if it passes signature and ZKP verification.
\end{enumerate}
Advantage is the success probability of producing a fresh valid tag.

\noindent
\emph{Proof Sketch:} Follows from EUF-CMA security of threshold BLS signatures and zk-SNARK soundness.

\paragraph{Property 4: Replay Resistance (RR).}
\label{prop:rr}
Any replay of $(\mathsf{chain\_id}, \mathsf{txid}, \mathsf{seq})$ is rejected; cross-domain replay is rejected via $\mathsf{chain\_id}$ binding (see Attack~\ref{attack:replay} for the corresponding adversarial goal; full proof in Appendix~\ref{P4}).

\noindent
\textbf{Game $\mathsf{Game}^{\mathsf{RR}}$:}
\begin{enumerate}[leftmargin=*]\setlength\itemsep{0em} 
    \item \emph{Setup:} $\mathcal{C}$ initializes the audit chain state.
    \item \emph{Challenge:} $\mathcal{A}$ attempts to submit a duplicate transaction record.
    \item \emph{Verify:} The audit chain rejects any duplicate.
\end{enumerate}

\noindent
\emph{Proof Sketch:} Sequence numbers and chain binding ensure uniqueness without relying on timestamps.

\paragraph{Property 5: IRP Privacy and Correctness.}
\label{prop:irp}
With fewer than $t$ shares, no coalition can learn $\mathsf{UID}$; with at least $t$ shares, reconstruction yields the correct $pk_{\mathsf{master}}$ Attack~\ref{attack:unauth-reveal} for the corresponding adversarial goal; full proof in Appendix~\ref{P5}).

\noindent
\textbf{Game $\mathsf{Game}^{\mathsf{IRP}}$:}
\begin{enumerate}[leftmargin=*]\setlength\itemsep{0em} 
    \item \emph{Setup:} $\mathcal{C}$ runs $(t,n)$-threshold encryption to produce $\mathsf{UID}$.
    \item \emph{Challenge:} $\mathcal{A}$ obtains $t'<t$ shares.
    \item \emph{Guess:} $\mathcal{A}$ outputs a guess for $\mathsf{UID}$.
\end{enumerate}
Advantage is $|\Pr[\text{correct guess}] - 1/|\mathcal{M}||$.

\noindent
\emph{Proof Sketch:} Directly from threshold ElGamal IND-CPA security.

\section{Evaluation}
\label{sec:Evaluation}

\subsection{Implementation and Parameter
Setup}

We implemented \VeilAudit as a prototype with three major components: 
(i) a transaction layer deployed on two independent EVM-compatible chains, 
(ii) an audit layer built with the Cosmos SDK, and 
(iii) a cross-chain bridge realized through Axelar’s General Message Passing (GMP).

The cryptographic components are implemented as follows: 
\AIP, \AUD, and \IRP are written in Solidity (Layer-1 contracts), Go (Ethereum), and Circom 2.2.2/SnarkJS 0.7.5 (ZK circuits). 
For zero-knowledge proofs we adopt Groth16 due to its mature ecosystem and compact proof size. We used about 20k lines of code to implement protocols and test tools.

Our experiments were conducted in two environments. 
\textbf{Local Workstation:} Ubuntu 22.04.5 LTS, Intel Xeon Gold 5220R @2.20GHz (48 cores), 376GB RAM, 16GB swap. 
\textbf{Cloud Deployment:} AWS, EC 20 nodes, Ubuntu 22.04,Intel Xeon Platinum 8269CY 4vCPU, 8GB RAM. The local workstation is primarily used for proof generation and verification, while the cloud Server supports distributed deployment of the cross-chain bridge and latency.

All Nodes and supporting services are containerized using Docker. 
Kubernetes is further employed to orchestrate node deployment and simulate distributed network topologies.

Unless otherwise specified, each reported data point is the mean of 10 independent runs, with 95\% confidence intervals computed via bootstrap resampling (1,000 resamples). We generate cross-chain transactions with a interaction patterns: escrow settlements (fund release after multi-party agreement). The cross-chain bridge is implemented via Axelar; For Auditor-Only Linkability (AOL), we construct ground-truth transaction pairs spanning multiple chains to measure equality-testing throughput, latency, and correctness. Details are given in Appendix~\ref{appendix:method}.


\subsection{Off-Chain Performance Evaluation}

This part validates the efficiency and practicality of \textit{VeilAudit} through micro-benchmarks on three key aspects. First, we demonstrate that all core off-chain cryptographic components operate in the \textbf{millisecond regime}. Second, we show that its zero-knowledge proofs require approximately \textbf{one second} for off-chain generation, while on-chain verification costs remain stable at \textbf{tens of milliseconds}. Finally, by analyzing the compact scale of our circuits, we substantiate the core design philosophy of being \emph{off-chain heavy, on-chain light}.

\noindent
\textbf{Performance of Core Cryptographic Primitives.} We first measure the latency of the cryptographic primitives used to derive anonymous addresses and prepare audit material. As shown in \textbf{Table~\ref{tab:primitives}}, all operations exhibit high efficiency. These \textbf{millisecond-scale} latency results indicate that \textit{VeilAudit}'s fundamental operations impose a negligible performance impact on the server and client-side, ensuring they do not become a system bottleneck.


\begin{table}[t]
  \centering
  \footnotesize
  \setlength{\tabcolsep}{3.5pt}
  \renewcommand{\arraystretch}{1.05}
  \rowcolors{2}{gray!5}{white}
  \begin{tabularx}{\columnwidth}{@{}%
      >{\raggedright\arraybackslash}l
      S[table-format=1.3]
      >{\raggedright\arraybackslash}X@{}}
    \rowcolor{gray!15}
    \textbf{Operation} & \textbf{Latency (ms)} & \textbf{Role in \textit{VeilAudit}} \\
    $\mathsf{AddrGen}$             & 1.069 & Derive a one-time anonymous address from the master key. \\
    $\mathsf{UID Enc}$            & 2.746 & Threshold-encrypt a masked identity for accountability. \\
    $\mathsf{CT}^{\text{link}}\ \mathsf{Enc}$   & 0.273 & Encrypt a linkage plaintext for equality testing. \\
$\mathsf{Commit}\!\big(\mathrm{Com}_{\mathrm{pk}}\big)$ & 1.307 & Commit to a public key for later ZK consistency. \\
     $m \gets H(\cdot)$    & 0.168 & Derive a global-domain message for digital signatures. \\
  \end{tabularx}
  \caption{Latency of core cryptographic primitives.}
  \label{tab:primitives}
\end{table}

\noindent
\textbf{Zero-Knowledge Circuit Performance and Scale.} VeilAudit relies on three Groth16 circuits: a control proof ($\pi_{\text{ctrl}}$), an execution proof ($\pi_{\text{exec}}$), and a consistency/linkage proof ($\pi_{\text{link}}$). We first analyze their R1CS scale, which forms the basis for their performance.

The experimental results validate our design goals. Owing to the compact circuit sizes shown in \textbf{Table~\ref{tab:zkp-scale-runtime}}, off-chain proving costs are stable at \textbf{1.1--1.2 seconds}, while on-chain verification latency is only \textbf{$\approx$20 milliseconds}. Combined with the millisecond-level latency of the core primitives, these results collectively demonstrate that \textit{VeilAudit}, as an \emph{off-chain heavy, on-chain light} system, offers a practical auditing solution that balances privacy and efficiency for high-throughput blockchain applications.


\begin{table}[t]
  \centering
  \footnotesize
  \setlength{\arrayrulewidth}{0pt}
  \renewcommand{\arraystretch}{1.08}
  \rowcolors{2}{gray!5}{white}
  \begin{tabularx}{\columnwidth}{|>{\raggedright\arraybackslash}X|
      S[table-format=2.3]|
      S[table-format=2.3]|
      S[table-format=2.3]|
      S[table-format=2.3]|
      S[table-format=4]}
    \hline
    \rowcolor{gray!15}
    \textbf{Circuit} & \textbf{Setup (s)} & \textbf{Witness (s)} &
    \textbf{Prove (s)} & \textbf{Verify (ms)} & \textbf{Constraints} \\
    \hline
    $\pi_{\text{ctrl}}$ & 2.064 & 0.196 & 1.159 & 18.927 & 1034 \\
    $\pi_{\text{exec}}$ & 1.690 & 0.194 & 1.132 & 18.834 &  517 \\
    $\pi_{\text{link}}$ & 2.192 & 0.203 & 1.226 & 19.151 & 1449 \\
    \hline
  \end{tabularx}
  \caption{Groth16 circuit scale and runtime for \textit{VeilAudit}.
  Setup is one-time; Witness/Prove are per-proof off-chain; Verify is
  the on-chain.}
  \label{tab:zkp-scale-runtime}
\end{table}

\subsection{On-Chain Gas Evaluation}

To quantify the on-chain footprint of \textit{VeilAudit}, we decompose a full transaction into three comparable scenarios:
\textbf{Baseline:} A standard public transaction consisting of an \textbf{ERC20 lock and unlock}. This scenario represents a typical on-chain interaction.
\textbf{VeilAudit (emit):} Builds upon the Baseline by adding the on-chain verification of three Groth16 proofs. The resulting audit tag is recorded as a low-cost log entry (an event), which is not persisted in the chain's state but is easily accessible to off-chain services.
\textbf{VeilAudit (store):} Also builds upon the Baseline with the same three proof verifications. However, the audit tag is persisted directly into the contract's state storage. This approach incurs a higher gas cost.

\textbf{Cost Analysis}
As summarized in \textbf{Table~\ref{tab:gas-3row}}, the on-chain overhead of \textit{VeilAudit} is dominated by ZK proof verification, which accounts for over \textbf{67\%} of the total gas overhead. This on-chain cost structure, combined with the computationally expensive proving step which incurs zero gas off-chain, validates our \emph{off-chain heavy, on-chain light} design. Furthermore, the audit tagging mechanism presents a significant cost trade-off, with event-based logging being substantially cheaper than storage persistence, offering developers architectural flexibility. \textit{VeilAudit} adds approximately \textbf{695k--872k gas} (\textasciitilde\$3.0--\$3.8 at 1 Gwei) on top of a baseline transaction. This cost is comparable to common DeFi interactions, indicating that \textit{VeilAudit} is economically viable for practical deployment.

\begin{table*}[t]
  \raggedright
  \small
  \setlength{\tabcolsep}{3pt}
  \renewcommand{\arraystretch}{1.04}
  \setlength{\arrayrulewidth}{0pt} 
  \rowcolors{2}{gray!5}{white}
  \begin{tabularx}{\textwidth}{%
    >{\raggedright\arraybackslash}X
    S[table-format=6.0]
    S[table-format=6.0]
    S[table-format=6.0]
    S[table-format=6.0]
    S[table-format=6.0]
    S[table-format=7.0]
    S[table-format=6.0]
    S[table-format=7.0]
    S[table-format=1.6]
    S[table-format=1.2]
  }
    \rowcolor{gray!15}
    \textbf{Scenario} & \textbf{Lock} & \textbf{Unlock} &
    {$\pi_{\text{ctrl}}$} & {$\pi_{\text{exec}}$} &
    {$\pi_{\text{link}}$} & {$\textbf{Total ZK}$} &
    \textbf{Atag} & \textbf{Total} & \textbf{ETH} & \textbf{USD} \\
    Baseline (lock+unlock) & 61967 & 54412 & 0 & 0 & 0 & 0 & 0 & 116379 & 0.000116 & 0.52 \\
    VeilAudit (emit)       & 61967 & 54412 & 221797 & 214585 & 228917 & 665299 & 29925  & 811603 & 0.000812 & 3.61 \\
    VeilAudit (store)      & 61967 & 54412 & 221797 & 214585 & 228917 & 665299 & 206533 & 988211 & 0.000988 & 4.39 \\
  \end{tabularx}
  \caption{End-to-end on-chain gas for baseline and \textit{VeilAudit} variants.
  Gasprice = 1 Gwei, 1 Ether = 109 Gwei, and 1 ETH=\$4{,}450.}
  \label{tab:gas-3row}
\end{table*}

\subsection{Auditor-Only Linkability (AOL): Equality and Clustering}

We evaluate the auditor-only linkability pipeline that operates on equality-test ciphertexts $\CTlink$. The goal is to show that (i) the equality stage sustains high throughput, and (ii) the downstream clustering faithfully reconstructs user groups.

\noindent
\textbf{Parameters setup.}
Each run processes a batch of $B$ audit tags (each with a $\CTlink$). The auditor’s \emph{visibility} is modeled by a sampling rate $p\in(0,1]$, i.e., the fraction of tags the auditor observes. We fix the number of latent users to $S$ and the mean tags-per-user to $\bar{k}$, and use two regimes to mimic low/high load:
\emph{low} $(B{=}10{,}000,~S{=}2{,}500,~\bar{k}{=}4)$ and
\emph{high} $(B{=}30{,}000,~S{=}7{,}500,~\bar{k}{=}4)$.
Unless otherwise noted, we use \texttt{repeats}$\,{=}5$ and \texttt{warmup}$\,{=}1$ per point.
Units are milliseconds (ms), seconds (s), and pairs per second (pairs/s).

\noindent
\textbf{Metrics.}
The equality stage performs trapdoor-enabled equality tests on candidate pairs derived from the visible set. The total pairs tested are \(N_{\mathrm{pairs}}\approx p\binom{B}{2}=p\,\frac{B(B-1)}{2}\). Let \(T_{\mathrm{batch}}\) be the wall-clock time for one batch (in seconds). We report equality throughput \(Q = N_{\mathrm{pairs}}/T_{\mathrm{batch}}\) (pairs/s). From the equality graph we recover clusters and compare to ground truth using two standard, unitless scores in \([0,1]\): Adjusted Rand Index (ARI) and Normalized Mutual Information (NMI); higher is better, and \(1.0\) indicates perfect recovery.

\begin{figure}[t]
  \centering
  \vspace{2pt}
  \begin{subfigure}{0.47\columnwidth} 
    \centering
    \includegraphics[width=\linewidth,trim=1 1 1 0,clip]{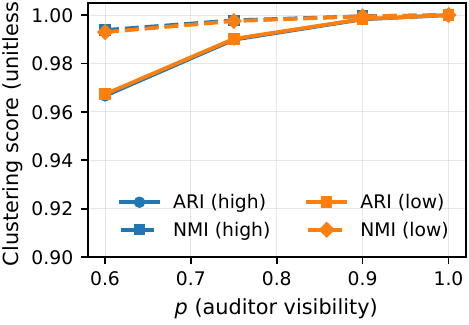}
    \caption{Accuracy vs.\ $p$}
    \label{fig:aol-accuracy}
  \end{subfigure}
  \hspace{0.03\columnwidth} 
  \begin{subfigure}{0.47\columnwidth}
    \centering
    \includegraphics[width=\linewidth,trim=1 1 1 0,clip]{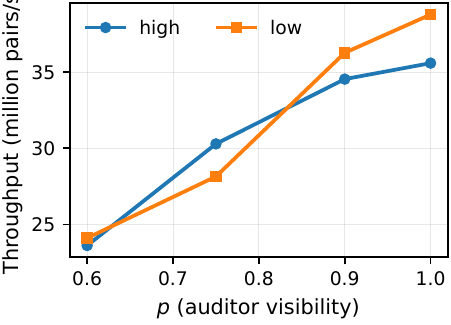}
    \caption{Throughput vs.\ $p$}
    \label{fig:aol-throughput}
  \end{subfigure}
  \vspace{1pt}

  \caption{AOL accuracy and throughput.}
  \label{fig:aol-both}
\end{figure}

\noindent
\textbf{Results.}
Across realistic auditor visibility (\(p\in[0.60,1.00]\)) and batch scales (\(\textit{low}: B{=}10\text{k},~\textit{high}: B{=}30\text{k}\)), the equality stage sustains tens of millions of comparisons per second without becoming a bottleneck. As shown in Fig.~\ref{fig:aol-throughput}, throughput remains in the \(\sim 24\text{–}39\)M pairs/s range: for \(B{=}10\text{k}\) we observe \(N_{\mathrm{pairs}}{=}30{,}004{,}980\) in \(1.245\,\mathrm{s}\) at \(p{=}0.60\) (\(\approx\!24\)M pairs/s) and \(49{,}995{,}000\) in \(1.289\,\mathrm{s}\) at \(p{=}1.00\) (\(\approx\!39\)M pairs/s); for \(B{=}30\text{k}\) the trend is similar (\(\sim 24\text{–}36\)M pairs/s). Clustering fidelity is already high at partial visibility and quickly approaches perfect: Fig.~\ref{fig:aol-accuracy} shows \(\mathrm{ARI}\approx 0.97\) and \(\mathrm{NMI}\approx 0.99\) at \(p{=}0.60\), rising to \(\approx 1.0\) by \(p\ge 0.90\). In short, \emph{VeilAudit} delivers engineering-grade scalability and auditing-grade accuracy (high ARI/NMI) under realistic settings.

\subsection{On-Chain Scalability}

We stress-test the on-chain step of \textit{VeilAudit} to answer two questions:
(i) does throughput scale linearly with the offered load, and
(ii) is end-to-end confirmation latency dominated by consensus cadence rather than contract logic?

\noindent
\textbf{Parameters setup.}
We benchmark two contract paths: \textbf{emitAtag} and \textbf{storeAtag}.
Blocks are produced every $T{=}500$\,ms.
For each run we fix the number of concurrent senders $N\!\in\!\{10,30,50,100\}$ and sweep the \emph{offered load} (QPS) $\mathcal{Q}\!\in\!\{5,10,20,40\}$.
Here, \textbf{QPS} means the aggregate \emph{submission rate} of transactions generated by clients (before inclusion); we enforce it with a client-side rate limiter that splits the target evenly across senders, so each sender emits $\mathcal{Q}/N$ tx/s (e.g., $\mathcal{Q}{=}20, N{=}10 \Rightarrow 2$ tx/s per sender).
Each operating point is sustained for $30$\,s, so the expected number of attempts is $\approx 30\mathcal{Q}$, which matches our “sent” counts.
We report \textbf{TPS} as the \emph{realized throughput} (mined tx/s) computed from receipts over the same window; when the mempool isn’t congested, TPS $\approx$ QPS, which is exactly what our results show.
Results are representative across $N$ because the dominant bound is the block cadence (not sender concurrency).

\noindent
\textbf{Metrics.}
P50/P95 are the 50/95th percentile of end-to-end confirmation time (submission~$\rightarrow$ inclusion), in milliseconds.
Average gas per transaction is reported for completeness.
Under fixed block interval $T$, a simple arrival model predicts median confirmation $\approx T/2$ and tail $\approx T$:
$\mathrm{P50}\approx T/2,\ \mathrm{P95}\approx T$
when contract execution is not bottleneck.

\begin{figure}[t]
  \centering
  \captionsetup{skip=2pt}
  \begin{subfigure}[t]{0.515\columnwidth}
    \centering
    \includegraphics[width=\linewidth]{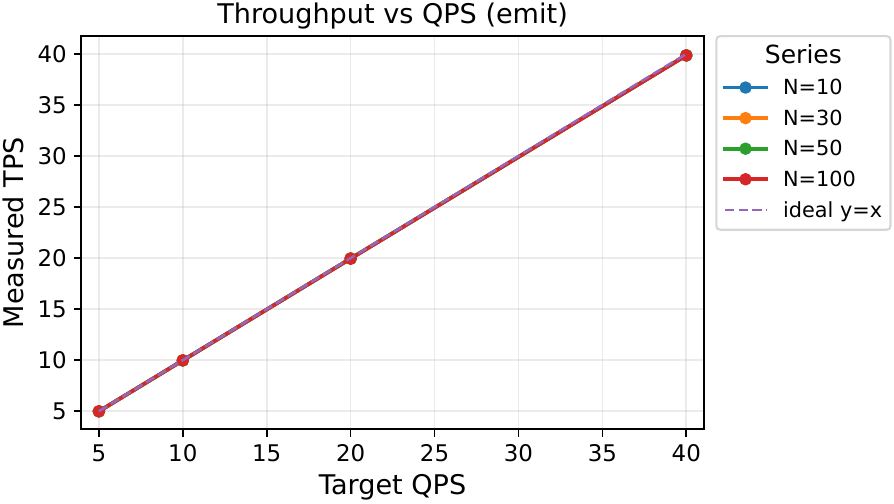}
    \caption{TPS vs.\ QPS (emit)}
    \label{fig:scal-tps:emit}
  \end{subfigure}\hspace{-0.03\columnwidth}
  \begin{subfigure}[t]{0.515\columnwidth}
    \centering
    \includegraphics[width=\linewidth]{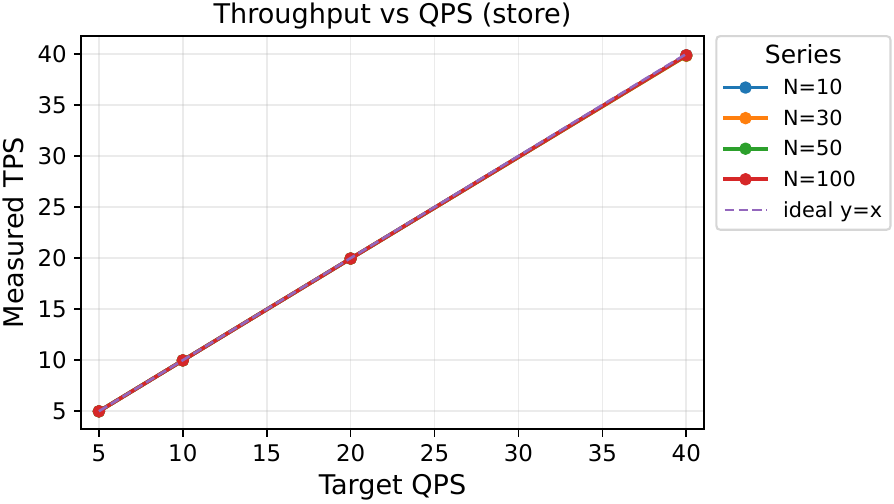}
    \caption{TPS vs.\ QPS (store)}
    \label{fig:scal-tps:store}
  \end{subfigure}
  \vspace{-0.2em}
  \caption{On-chain throughput under 500\,ms blocks.}
  \label{fig:scal-tps}
\end{figure}

\begin{figure}[t]
  \centering
  \begin{subfigure}[t]{0.495\columnwidth}
    \centering
    \includegraphics[width=\linewidth]{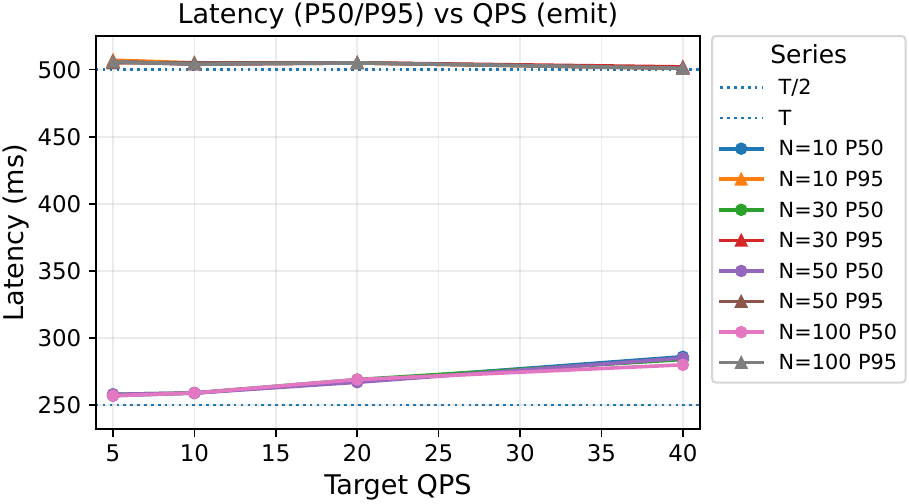}
    \caption{P50/P95 latency vs.\ QPS (emit)}
    \label{fig:scal-lat:emit}
  \end{subfigure}\hfill
  \begin{subfigure}[t]{0.495\columnwidth}
    \centering
    \includegraphics[width=\linewidth]{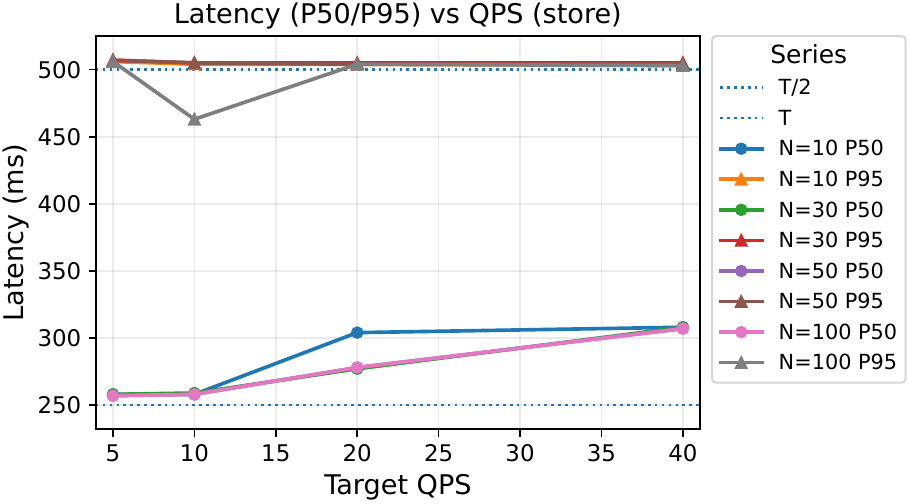}
    \caption{P50/P95 latency vs.\ QPS (store)}
    \label{fig:scal-lat:store}
  \end{subfigure}
  \caption{End-to-end latency under 500\,ms blocks.}
  \label{fig:scal-lat}
\end{figure}

\noindent
\textbf{Results.}
Across offered loads $Q\!\in\!\{5,10,20,40\}$, realized throughput tracks the ideal line ($\mathrm{TPS}\!\approx\!Q$) in both recording modes—see Fig.~\ref{fig:scal-tps}\subref{fig:scal-tps:emit} (\texttt{emit}) and Fig.~\ref{fig:scal-tps}\subref{fig:scal-tps:store} (\texttt{store}). Confirmation latency is governed by the 500\,ms block cadence rather than contract execution: median $\mathrm{P50}\!\approx\!T/2$ and tail $\mathrm{P95}\!\approx\!T$ with $T{=}500$\,ms in both modes (Fig.~\ref{fig:scal-lat}\subref{fig:scal-lat:emit}, \subref{fig:scal-lat:store}), and is stable across $N\!\in\!\{10,30,50,100\}$. In short, \textit{VeilAudit} preserves line-rate scalability and adds negligible latency under realistic loads.

\section{Related Work}
\label{sec:related}

We review four lines of research related to \textsc{VeilAudit}: (i) trust-minimized cross-chain communication and proof-based bridges, (ii) zero-knowledge proofs that enable auditable privacy on public ledgers, (iii) deanonymization and graph-based analytics from the auditor's perspective, and (iv) threshold cryptography for regulated disclosure.

\noindent
\textbf{Trustless Cross-Chain Communication and Proof-Based Bridges.}
A large body of work investigates \emph{trust-minimized} cross-chain messaging using succinct clients and proof-verified state transitions, thereby avoiding trusted relays and reducing verification costs on destination chains~\cite{xie2022zkbridge,zamyatin2019xclaim,guo2024zkcross,narula2018zkledger,corrigan2017prio}. These systems primarily optimize \emph{executability correctness} of asset movement or message passing across heterogeneous chains. In contrast, \textsc{VeilAudit} targets \emph{auditor-facing verifiability}: our commit--prove--record workflow yields compact \emph{audit tags} consumable by an audit chain, providing publicly verifiable evidence of correct execution and linkage while keeping identities encrypted for policy-gated access.

\noindent
\textbf{ZK for Auditable Privacy on Public Ledgers.}
ZK systems reconcile transparency with confidentiality via succinct on-chain verification~\cite{sasson2014zerocash,groth2016size,bunz2018bulletproofs,bunz2020flyclient,zamyatin2019xclaim,zamyatin2021sok}. These designs inform our circuit choices (binding/hiding commitments, link proofs, and amortized verification). However, most do not expose a \emph{behavior-linkage without identity} interface for auditors, nor a first-class, policy-bound threshold decryption path. \textsc{VeilAudit} explicitly separates \emph{behavioral linkage} (via commitments and a link proof) from \emph{identity revelation} (via threshold decryption of an encrypted identity capsule), so routine audits never require deanonymization.

\noindent
\textbf{Deanonymization, Clustering, and Auditor-Side Attacks.}
Prior studies show that timing, denomination patterns, and multi-hop flows can enable re-identification via clustering on transaction graphs~\cite{meiklejohn2013fistful,kalodner2020blocksci,chen2022survey}. 
We treat them as the adversarial capability our design aims to neutralize. 
\textsc{VeilAudit} deliberately withholds the features those attacks rely on: plaintext addresses and linkable identifiers are never exposed; amounts/denominations are hidden; cross-tag linkage is revealed only through trapdoor-gated equality proofs; and identities remain threshold-encrypted. 
Our goal is that, absent authorized decryption shares, an honest-but-curious auditor gains no advantage beyond protocol-visible events. 
Accordingly, our experiments evaluate linkage recoverability \emph{under the VeilAudit interface} (AOL equality + clustering on audit tags), rather than reproducing public-graph deanonymization heuristics.

\noindent
\textbf{Threshold Cryptography and Regulated Disclosure.}
Threshold encryption and verifiable secret sharing align accountability with confidentiality~\cite{shamir1979share,feldman1987practical,desmedt1994homomorphic,atapoor2023vss,gennaro2018fast}. \textsc{VeilAudit} integrates a $t$-out-of-$n$ decryption path directly into the protocol semantics: audit tags carry threshold-encrypted identity capsules; policy compliance is evidenced on-chain; and only when a regulator quorum attests to policy predicates are decryption shares combined. Our analysis covers confidentiality against $<t$ colluding authorities and correctness of share verification/combination, yielding compliant identity revelation with cryptographic auditability.

\section{Conclusion}
\label{sec:conclusion}

This paper addressed the fundamental tension between privacy and accountability in cross-chain systems. We introduced \textsc{VeilAudit}, a framework that breaks this deadlock through a novel primitive: Auditor-Only Linkability (AOL). Our core mechanism, the Linkable Audit Tag, combines zero-knowledge proofs with specialized encryption to allow auditors to trace anonymous behaviors without deanonymizing users, while a threshold-gated protocol ensures accountability is possible only under due process.

Our prototype and evaluation demonstrate that \textsc{VeilAudit} is practical: it adds a modest on-chain footprint comparable to common DeFi interactions, while the off-chain auditing pipeline achieves engineering-grade scalability and accuracy. By enabling a trustworthy pseudonymous economy, \textsc{VeilAudit} provides a foundational layer for balancing capital efficiency with compliance in the decentralized future.

\cleardoublepage
\bibliographystyle{plain}   

\cleardoublepage
\appendix
\section*{Appendix}

\section{Preliminaries}
\label{sec:Preliminaries}

This section introduces the core cryptographic and blockchain foundations used throughout the design of VeilAudit, including public-key wallet models, zk-SNARKs, cross-chain bridges, and threshold encryption. These primitives collectively enable privacy-preserving yet auditable interactions across multiple heterogeneous blockchain environments.

\subsection{Public-Key Wallets}

Public-key wallets are the foundational identity mechanism in most blockchain systems. Users independently generate a private key $sk \in \mathbb{Z}_q$, where $q$ is a large prime defining the order of an elliptic curve group $\mathbb{G}$. The corresponding public key is computed as $pk = sk \cdot G$, where $G$ is the standard generator of $\mathbb{G}$.

In Ethereum and many EVM-compatible blockchains, the account address is derived from the public key via a hash-based procedure: 
$\mathsf{addr} = \mathsf{LSB}_{160}(\mathsf{Keccak256}(pk))$, 
where $\mathsf{LSB}_{160}$ denotes the least-significant 160 bits extraction. This ensures the address is compact, efficiently verifiable, and collision-resistant under standard cryptographic assumptions. For any fixed $sk$, the resulting $(pk, \mathsf{addr})$ pair is uniquely determined. Given that $sk$ is chosen uniformly at random from $\mathbb{Z}_q$, the entropy of the key space is approximately $2^{256}$, making key collision or impersonation attacks computationally infeasible.

In \textsc{VeilAudit}, we explicitly retain this standard wallet structure without requiring any key registration, identity binding, or modification of the existing account generation logic. All cryptographic derivations are constructed on top of the native wallet model, ensuring compatibility with existing infrastructures while preserving decentralization and self-sovereign identity.

\subsection{zk-SNARKs}

Succinct Non-Interactive Arguments of Knowledge (zk-SNARKs) are cryptographic proof systems that enable a prover to convince a verifier that a certain computation was correctly performed, without revealing any sensitive input data.

Formally, a zk-SNARK proof system is defined as a tuple $(\mathsf{Gen}, \mathsf{Prove}, \mathsf{Verify})$:
\begin{itemize}\setlength\itemsep{0em} 
\item $\mathsf{Gen}(C) \rightarrow (\mathsf{pk}, \mathsf{vk})$: Given a computation circuit $C$, a trusted setup phase generates a proving key $\mathsf{pk}$ and a verification key $\mathsf{vk}$.
\item $\mathsf{Prove}(\mathsf{pk}, x, w) \rightarrow \pi$: A prover holding a witness $w$ for input $x$ produces a succinct proof $\pi$ attesting that $(x, w) \in R_C$.
\item $\mathsf{Verify}(\mathsf{vk}, x, \pi) \rightarrow \{0,1\}$: Anyone with the verification key $\mathsf{vk}$ can check the validity of $\pi$ on input $x$.
\end{itemize}

In \textsc{VeilAudit}, zk-SNARKs are used to achieve two core guarantees. First, they prove that an anonymous audit tag is correctly derived from a legitimate secret key and contextual parameters, without disclosing the key or linking it to any identity. Second, zk-SNARKs ensure that the contents embedded in the tag—such as transaction metadata relevant for audit—are correctly constructed according to a well-formed circuit, enabling consistent validation across chains. Both usages rely on non-interactive proof generation and public verification, ensuring scalability and trustless auditability.

\subsection{Cross-Chain Bridges}
\label{sec:crosschain-bridges}

A \emph{cross-chain bridge} is a protocol that enables the secure transfer of data or assets between heterogeneous blockchains. 
Bridges typically operate under one of several architectural paradigms: \emph{notary schemes}, where a quorum of trusted parties attest to cross-chain events; \emph{hashed time-lock contracts} (HTLCs), where conditional payments are enforced via hash preimages and timeouts; or \emph{light-client–based relays}, where block headers and Merkle proofs are verified across chains to ensure event authenticity.

In \textsc{VeilAudit}, we instantiate the bridge layer using the \textbf{Axelar network}, a decentralized interoperability platform that supports message passing across multiple EVM-compatible chains and Cosmos SDK–based chains. 
Axelar's consensus employs a \emph{threshold signature scheme} (TSS) to collectively sign outbound messages, ensuring that no single relayer can forge cross-chain data. 
Each cross-chain message is subject to a \emph{confirmation depth} policy, requiring that the source-chain transaction is finalized with a sufficient number of confirmations before relay, thereby mitigating reorganization attacks.
Furthermore, Axelar implements \emph{deduplication} on the receiving chain to prevent replay of previously delivered messages, which is critical for the correctness of audit-tag provenance.

\subsection{Threshold Encryption}

Threshold encryption is a cryptographic primitive that enables secret recovery only when a quorum of authorized parties collaborate. Specifically, a secret \(m\) is encrypted under a public key \(pk\) such that it can only be decrypted when at least \(t\) out of \(n\) designated decryption parties participate. This technique enhances robustness, accountability, and resistance to abuse in sensitive operations such as identity de-anonymization.

In the context of \textsc{VeilAudit}, threshold encryption serves a critical role in supporting \emph{authorized identity revelation}. Here, a user’s identity is defined as their \textbf{original wallet public key} \(pk_{\mathsf{master}}\)—prior to any salt-based masking or pseudonymous derivation. This value remains hidden under a threshold-encrypted payload during normal operation.

To formalize, let:
\begin{itemize}\setlength\itemsep{0em} 
\item \(\mathsf{Enc}_{\tpk}(pk_{\mathsf{master}})\) be the ciphertext encrypted under a threshold public key \(\tpk\).
\item \(\mathsf{Dec}_{\mathsf{sk}_i}\) denote the partial decryption share from authority \(i\).
\end{itemize}

The de-anonymization process is triggered only upon satisfying the on-chain authorization policy \(\mathsf{Policy}_\mathsf{reveal}\) recorded on the audit chain. Once the number of valid signatures reaches the predefined threshold \(t\), the original identity is reconstructed:
\[
\left\{ \mathsf{Dec}_{\mathsf{sk}_1}, \ldots, \mathsf{Dec}_{\mathsf{sk}_t} \right\} \xRightarrow[]{\text{Reconstruction}} pk_{\mathsf{master}}
\]
\[
\mathsf{Reveal}(\mathsf{uid}_i) \rightarrow pk_{\mathsf{master}} \text{ only if } \left| \Sigma_{\mathsf{reveal}} \right| \geq t
\]
Here, \(\left| \Sigma_{\mathsf{reveal}} \right|\) denotes the number of \emph{valid authorization signatures} collected from distinct approved parties.

This mechanism guarantees that no single party can reveal a user's true identity unilaterally. It also ensures \textbf{auditability} of the de-anonymization process itself: each authorization event is publicly recorded, and every signature is verifiable against a registered public key.

By treating identity as an encrypted commitment and requiring multiparty consent for its disclosure, \textsc{VeilAudit} balances privacy protection with regulatory accountability in cross-chain transaction auditing.

\subsection{Public-Key Encryption with Equality Test (PKE-ET)}
\label{sec:pkeet}

\emph{Public-Key Encryption with Equality Test} (PKE-ET)~\cite{zhang2008pkeet} is a cryptographic primitive that extends conventional public-key encryption to support equality comparison on ciphertexts without revealing the underlying plaintext. 
In addition to the usual public/secret key pair $(pk, sk)$, a PKE-ET system generates a separate \emph{equality-test key} $\mathsf{etk}$ that enables a designated tester to determine whether two ciphertexts encrypt the same message, while learning nothing about the message itself.

A PKE-ET scheme is defined by the following algorithms:
\begin{itemize}\setlength\itemsep{0em}
    \item $\mathsf{KeyGen}(1^\lambda) \rightarrow (pk, sk, \mathsf{etk})$: On input the security parameter $\lambda$, output a public key $pk$, a secret key $sk$, and an equality-test key $\mathsf{etk}$.
    \item $\mathsf{Enc}(pk, m) \rightarrow c$: Encrypt a message $m$ under $pk$ to obtain ciphertext $c$.
    \item $\mathsf{Dec}(sk, c) \rightarrow m$: Decrypt ciphertext $c$ under $sk$ to recover the plaintext $m$.
    \item $\mathsf{Test}(\mathsf{etk}, c_1, c_2) \rightarrow \{0,1\}$: Using $\mathsf{etk}$, output $1$ if $c_1$ and $c_2$ encrypt the same message, and $0$ otherwise.
\end{itemize}

In \textsc{VeilAudit}, the plaintext $m$ encrypted under PKE-ET is a \emph{global behavior-equivalence string}: 
$m = \mathsf{Hash}(\Com_{\texttt{pk}} \parallel \textsf{domain})$, 
where $\Com_{\texttt{pk}}$ is a commitment to the user's master public key and $\textsf{domain}$ is a contextual domain separator (e.g., chain identifier).  
The corresponding ciphertext $\mathsf{CT}^{\text{link}}$ is included in each $\texttt{Atag}$, allowing an auditor---who possesses $\mathsf{etk}$---to determine whether two tags originate from the same committed identity without learning the identity itself.  
The public, lacking $\mathsf{etk}$, cannot perform any such linkage, thus preserving unlinkability.

\section{Algorithm}
\begin{algorithm}[H]
\caption{Anonymous Identity Protocol (\(\mathcal{P}_{\text{AIP}}\))}
\label{alg:AIP}
\begin{algorithmic}[1]
\Require Long-term master keypair \((\texttt{sk}_{\text{master}}, \texttt{pk}_{\text{master}})\);
         threshold public key \(\texttt{tpk}\);
         equality-test public key \(\texttt{apk}\);
         escrow contracts \(\texttt{Esc}_i\) (per chain);
         domain tags \(\texttt{ctx}_{\textsf{CTRL}}, \texttt{ctx}_{\textsf{EXEC}}, \texttt{ctx}_{\textsf{GLOBAL}}\).
\Ensure Tuple
  \((\texttt{addr}_i^{\text{anon}},
    \pi_{\text{ctrl}}^i,
    \pi_{\text{exec}},
    \texttt{UID}_i,
    \texttt{Com}_{\texttt{pk},i},
    \CTlink_i,
    \pi_{\text{link}}^i)\).

\State \textbf{(Ephemeral address)} Sample fresh salt \(\texttt{salt}_{\text{addr}}\).
\State Derive \(\texttt{sk}_i^{\text{anon}} \leftarrow
  \mathsf{KDF}(\texttt{sk}_{\text{master}}, \texttt{salt}_{\text{addr}})\).
\State Compute \(\texttt{pk}_i^{\text{anon}} \leftarrow
  \texttt{sk}_i^{\text{anon}} \cdot G\),
  \(\texttt{addr}_i^{\text{anon}} \leftarrow \mathsf{Addr}(\texttt{pk}_i^{\text{anon}})\).

\State \textbf{(Control proof)} Let
  \(c_{\textsf{CTRL}} \leftarrow
   H(\texttt{ctx}_{\textsf{CTRL}} \parallel
     \texttt{addr}_i^{\text{anon}} \parallel
     \texttt{nonce}_{\textsf{sess}})\).
\State Produce \(\pi_{\text{ctrl}}^i\) using Circuit~\ref{circuit:ctrl}
  with challenge \(c_{\textsf{CTRL}}\).
\State Submit \(\pi_{\text{ctrl}}^i\) to \(\texttt{Esc}_i\);
  upon on-chain verification, \(\texttt{Esc}_i\) funds
  \(\texttt{addr}_i^{\text{anon}}\).

\State \textbf{(Execute x-chain)} Perform cross-chain transfer
  between anonymous addresses (coordinated by bridge).
\State Obtain bridge attestation \(\sigma_{\textsf{Axelar}}\)
  and message \(m_{\textsf{exec}}\) (includes
  \(\texttt{src}, \texttt{dst}, \texttt{txid}, \texttt{nonce},
   \texttt{depth}, \ldots\)).

\State \textbf{(Exec proof)} Compute domain-separated statement
  \(c_{\textsf{EXEC}} \leftarrow
   H(\texttt{ctx}_{\textsf{EXEC}} \parallel m_{\textsf{exec}})\).
\State Produce \(\pi_{\text{exec}}\) using Circuit~\ref{circuit:exec}
  (verifies Axelar TSS signature and binds anti-replay fields).
\State Post \(\pi_{\text{exec}}\) to the audit pipeline.

\State \textbf{(Identity encryption)} Set
  \(\texttt{UID}_i \leftarrow \mathsf{Enc}_{\texttt{tpk}}(\texttt{pk}_i)\).

\State \textbf{(Commitment \& ET)} Let
  \(x \leftarrow H_{\mathbb{F}}(\texttt{pk}_i)\),
  sample fixed per-user \(r_\star\) (long-term),
  and per-transaction \(r\) (fresh).
\State Compute \(\texttt{Com}_{\texttt{pk},i} \leftarrow g^x h^{r_\star}\).
\State Define \(m \leftarrow
  H(\texttt{Com}_{\texttt{pk},i} \parallel \texttt{ctx}_{\textsf{GLOBAL}})\).
\State Compute \(\CTlink_i \leftarrow \mathsf{ET.Enc}(\texttt{apk}, m; r)\).

\State \textbf{(Consistency proof)} Produce \(\pi_{\text{link}}^i\)
  with Circuit~\ref{circuit:link}, proving consistency among
  \(\texttt{UID}_i, \texttt{Com}_{\texttt{pk},i}, \CTlink_i\).

\State \Return
  \((\texttt{addr}_i^{\text{anon}},
    \pi_{\text{ctrl}}^i,
    \pi_{\text{exec}},
    \texttt{UID}_i,
    \texttt{Com}_{\texttt{pk},i},
    \CTlink_i,
    \pi_{\text{link}}^i)\).
\end{algorithmic}
\end{algorithm}

\begin{algorithm}[H]
\caption{Audit Protocol (\(\mathcal{P}_{\text{AUD}}\))}
\label{alg:PAUD}
\begin{algorithmic}[1]
\raggedright
\Require Source/destination chain ids \(\mathsf{cid}_{\mathrm{src}}, \mathsf{cid}_{\mathrm{dst}}\);
         txids \(\mathsf{txid}_{\mathrm{src}}, \mathsf{txid}_{\mathrm{dst}}\);
         bridge message id \(\mathsf{msgid}\);
         timestamp \(\mathsf{ts}\);
         finality depth \(\Delta\);
         execution proof \(\pi_{\text{exec}}\).
\Require For \(i \in \{A,B\}\): \((\texttt{UID}_i,\ \texttt{Com}_{\texttt{pk},i},\ \CTlink_i,\ \pi_{\text{link}}^i)\).
\Require Domain tags \(\texttt{ctx}_{\textsf{EXEC}},\ \texttt{ctx}_{\textsf{GLOBAL}}\).
\Ensure Immutable audit record on the audit chain (or rejection on failure).

\State \textbf{(Dedup key)} \(k \leftarrow H(\mathsf{cid}_{\mathrm{dst}} \parallel \mathsf{txid}_{\mathrm{dst}} \parallel \mathsf{msgid})\).
\State \textbf{require} \(\neg\,\textsc{Exists}(k)\). \Comment{replay/dedup guard}

\State \textbf{(Verify execution)} \textbf{require } \(\textsc{VerifyExec}(\pi_{\text{exec}})=1\).
\Statex binds: \(\mathsf{cid}_{\mathrm{src}}, \mathsf{txid}_{\mathrm{src}}, \mathsf{cid}_{\mathrm{dst}}\).
\Statex and \(\mathsf{txid}_{\mathrm{dst}}, \mathsf{msgid}, \Delta, \texttt{ctx}_{\textsf{EXEC}}\).

\State \textbf{(Verify link for parties)} For \(i \in \{A,B\}\) \textbf{require } \(\textsc{VerifyLink}(\pi_{\text{link}}^i)=1\).
\Statex binds: \(\texttt{UID}_i, \texttt{Com}_{\texttt{pk},i}, \CTlink_i, \texttt{ctx}_{\textsf{GLOBAL}}\).

\State \textbf{(Assemble core)} \(\texttt{Atag.core} \leftarrow (\mathsf{cid}_{\mathrm{src}}, \mathsf{txid}_{\mathrm{src}}, \mathsf{cid}_{\mathrm{dst}}, \mathsf{txid}_{\mathrm{dst}}, \mathsf{msgid}, \mathsf{ts})\).

\State \textbf{(Bundle A)} \(\texttt{Atag.A} \leftarrow (\texttt{UID}_A, \texttt{Com}_{\texttt{pk},A}, \CTlink_A, \pi_{\text{link}}^A)\).
\State \textbf{(Bundle B)} \(\texttt{Atag.B} \leftarrow (\texttt{UID}_B, \texttt{Com}_{\texttt{pk},B}, \CTlink_B, \pi_{\text{link}}^B)\).

\State \textbf{(Aggregate)} \(\texttt{Atag} \leftarrow (\texttt{Atag.core};\ \pi_{\text{exec}};\ \texttt{Atag.A};\ \texttt{Atag.B})\).

\State \textbf{(Commit)} \(\textsc{Store}(k,\ \texttt{Atag})\);
       \(\textsc{Emit}\ \textsf{TagCommitted}(k,\ H(\texttt{Atag}))\).
\State \Return \(k\).
\end{algorithmic}
\end{algorithm}

\begin{algorithm}[H]
\caption{Identity Revealed Protocol ($\mathcal{P}_{\text{IRP}}$)}
\label{alg:IRP}
\begin{algorithmic}[1]
\Require Case identifier \texttt{caseID};
\Statex \quad Implicated tag set $\mathcal{S}=\{\texttt{Atag}_1,\dots,\texttt{Atag}_k\}$;
\Statex \quad Regulator policy $\Pi_{\text{reg}}$;
\Statex \quad Committee $\mathbb{G}=\{\mathcal{G}_1,\dots,\mathcal{G}_n\}$ with threshold $t$.
\Ensure Recovered master public keys $\{\texttt{pk}^{\text{master}}_u\}$ (possibly empty).

\State \textbf{Request.} Auditor submits
$\langle \texttt{caseID},\, \mathcal{S},\, \text{clusterEvidence} \rangle$.

\State \textbf{Vote.} Each $\mathcal{G}_j \in \mathbb{G}$ issues
\textsf{approve}/\textsf{deny} according to $\Pi_{\text{reg}}$.

\If{$\#\{\textsf{approve}\} < t$}
  \State \Return $\varnothing$ \Comment{insufficient approvals}
\EndIf

\State Record authorization artifact:
\Statex \quad approving set, timestamp.

\State \textbf{UID aggregation.} Extract
$\mathcal{U} \gets \{\, \texttt{UID} \mid
\texttt{UID} \in \texttt{Atag}_\ell,\ \texttt{Atag}_\ell \in \mathcal{S}\,\}$.

\ForAll{$\texttt{UID} \in \mathcal{U}$}
  \State \textbf{Shares.} Collect partial decryptions:
  \Statex \quad
  $\mathcal{D} \gets \{\mathsf{Dec}_{\mathcal{G}_j}(\texttt{UID})\}_{j \in \mathcal{A}}$,
  with $|\mathcal{A}|\ge t$.
  \State \textbf{Combine.} Compute
  $\texttt{pk}^{\text{master}} \gets \mathsf{Combine}(\mathcal{D})$.
  \State Append $\texttt{pk}^{\text{master}}$ to the output set.
\EndFor

\State \textbf{Materialize.} Emit on the audit chain:
\Statex \quad
\textsf{IdentityRevealed}(
\texttt{caseID},
$\mathcal{S}$,
authorization artifact,
$\{\texttt{pk}^{\text{master}}\}$).

\State \Return $\{\texttt{pk}^{\text{master}}\}$.
\end{algorithmic}
\end{algorithm}

\section{ZK Circuit}

\begin{algorithm}[H]
\caption{ZK Circuit: CTRL (\(\pi_{\text{ctrl}}^i\)) --- Control of Anonymous Address}
\label{circuit:ctrl}
\begin{algorithmic}[1]
\Require \textbf{Public Inputs:}
  anonymous address \(\texttt{addr}_i^{\text{anon}}\);
  challenge \(c_{\textsf{CTRL}}\).
\Require \textbf{Private Witness:}
  \(\texttt{sk}_i^{\text{anon}}\),
  signature \(\sigma_{\textsf{anon}}\) on \(c_{\textsf{CTRL}}\).
\Ensure \textbf{Public Outputs:} none.

\State \textbf{Key relation:}
  reconstruct \(\texttt{pk}_i^{\text{anon}} \leftarrow
  \texttt{sk}_i^{\text{anon}} \cdot G\).
\State \textbf{Address check:}
  enforce \(\mathsf{Addr}(\texttt{pk}_i^{\text{anon}})
  = \texttt{addr}_i^{\text{anon}}\).
\State \textbf{Signature check:}
  enforce \(\mathsf{VerifySig}(c_{\textsf{CTRL}},
    \sigma_{\textsf{anon}}, \texttt{pk}_i^{\text{anon}}) = 1\).
\State \textbf{Domain separation:}
  \(c_{\textsf{CTRL}}\) is hashed with \(\texttt{ctx}_{\textsf{CTRL}}\),
  binding session nonce and address.
\State \textbf{Soundness:}
  the circuit proves knowledge of \(\texttt{sk}_i^{\text{anon}}\)
  that controls \(\texttt{addr}_i^{\text{anon}}\), without revealing it.
\end{algorithmic}
\end{algorithm}

\begin{algorithm}[H]
\caption{ZK Circuit: EXEC (\(\pi_{\text{exec}}\)) --- Cross-Chain Execution Validity}
\label{circuit:exec}
\begin{algorithmic}[1]
\Require \textbf{Public Inputs:}
  message summary \(m_{\textsf{exec}}\) (includes
  \(\texttt{src}, \texttt{dst}, \texttt{txid}, \texttt{nonce}, \texttt{depth}\));
  Axelar threshold public key \(\texttt{tpk}_{\textsf{Axelar}}\).
\Require \textbf{Private Witness:}
  bridge signature \(\sigma_{\textsf{Axelar}}\);
  optional Merkle proof(s) of inclusion for
  attestation batch or gateway state.
\Ensure \textbf{Public Outputs:}
  anti-replay nullifier \(\mathsf{null}\).

\State \textbf{Domain binding:}
  compute \(c_{\textsf{EXEC}} \leftarrow
   H(\texttt{ctx}_{\textsf{EXEC}} \parallel m_{\textsf{exec}})\).
\State \textbf{Signature check:}
  enforce \(\mathsf{VerifyTSS}_{\textsf{Axelar}}
  (c_{\textsf{EXEC}}, \sigma_{\textsf{Axelar}},
   \texttt{tpk}_{\textsf{Axelar}}) = 1\).
\State \textbf{Inclusion (optional):}
  if batching is used, verify Merkle inclusion of
  \((c_{\textsf{EXEC}}, \sigma_{\textsf{Axelar}})\)
  in the published batch root.
\State \textbf{Anti-replay:}
  compute \(\mathsf{null} \leftarrow
  H(\texttt{txid} \parallel \texttt{nonce} \parallel
    \texttt{src} \parallel \texttt{dst})\).
\State \textbf{Output:}
  expose \(\mathsf{null}\) as public output so that
  the audit contract can mark it used exactly once.
\end{algorithmic}
\end{algorithm}

\begin{algorithm}[H]
\caption{ZK Circuit: LINK (\(\pi_{\text{link}}^i\)) --- Consistency of \(\texttt{UID}_i\), \(\texttt{Com}_{\texttt{pk},i}\), \(\CTlink_i\)}
\label{circuit:link}
\begin{algorithmic}[1]
\Require \textbf{Public Inputs:}
  \(\texttt{UID}_i\);
  \(\texttt{Com}_{\texttt{pk},i}\);
  \(\CTlink_i\);
  audit domain \(\texttt{ctx}_{\textsf{GLOBAL}}\).
\Require \textbf{Private Witness:}
  \(\texttt{pk}_i\), \(x \leftarrow H_{\mathbb{F}}(\texttt{pk}_i)\);
  long-term randomness \(r_\star\);
  ET randomness \(r\).
\Ensure \textbf{Public Outputs:} optional \(\texttt{Com}_m\) (binding).

\State \textbf{Commitment validity:}
  enforce \(\texttt{Com}_{\texttt{pk},i} = g^x h^{r_\star}\)
  with \(x = H_{\mathbb{F}}(\texttt{pk}_i)\).
\State \textbf{UID correctness:}
  enforce \(\texttt{UID}_i = \mathsf{Enc}_{\texttt{tpk}}(\texttt{pk}_i)\)
  via an \emph{encryption correctness} relation
  (scheme-specific constraint).
\State \textbf{Behavioral plaintext:}
  compute \(m \leftarrow
  H(\texttt{Com}_{\texttt{pk},i} \parallel \texttt{ctx}_{\textsf{GLOBAL}})\).
\State \textbf{ET-ciphertext validity:}
  enforce \(\CTlink_i = \mathsf{ET.Enc}(\texttt{apk}, m; r)\)
  (scheme-specific constraint).
\State \textbf{Bind output (optional):}
  set \(\texttt{Com}_m \leftarrow \mathsf{Com}(m)\)
  and expose it as a public output if the implementation
  uses external binding of \(m\).
\State \textbf{Privacy note:}
  the circuit reveals none of
  \(\texttt{pk}_i, r_\star, r\),
  yet proves all three artifacts are consistent w.r.t.\ the
  same hidden identity.
\end{algorithmic}
\end{algorithm}

\section{Formal Proofs for Security Properties}
\label{app:proofs}

This appendix provides full game-based reductions for the properties stated in Section~\ref{sec:security}, under the adversary capabilities specified in the Threat Model (Section~\ref{sec:threatmodel}). We consider PPT adversaries and work in the Random Oracle Model (ROM) for the hash $H$.

\subsection{Preliminaries and Notation}

\paragraph{Assumptions.}
(i) zk-SNARK $\Pi_{\mathsf{SN}}$ is complete, sound, and zero-knowledge; 
(ii) Pedersen commitments are perfectly hiding and computationally binding; 
(iii) Threshold ElGamal is IND-CPA secure in the $t$-out-of-$n$ setting; 
(iv) Public-key encryption with equality test (PKE-ET) is semantically secure and supports equality testing \emph{only} with the equality-test key; 
(v) BLS threshold signatures are EUF-CMA secure.

\paragraph{Notation (as in the main text).}
$\pk_{\mathsf{master}}$: user’s master public key; 
$\mathsf{Com}_{\pk}$: Pedersen commitment to $\pk_{\mathsf{master}}$; 
$\mathsf{UID}$: threshold-ElGamal ciphertext of $\pk_{\mathsf{master}}$; 
$\mathsf{CT}^{\mathsf{link}}$: PKE-ET ciphertext of $m := H(\mathsf{Com}_{\pk} \parallel \mathsf{ctx}_{\mathsf{GLOBAL}})$; 
$\pi_{\mathsf{link}}$: zk proof that $(\mathsf{Com}_{\pk}, \mathsf{UID}, \mathsf{CT}^{\mathsf{link}})$ are consistently derived per the circuit; 
$\texttt{Atag}$: audit tag containing the above and provenance fields $(\mathsf{chain\_id}, \mathsf{txid}, \mathsf{seq})$ plus the Axelar-side threshold signature.

\subsection{Property~\ref{prop:mda}: Minimal Disclosure Auditing}
\label{P1}

\paragraph{Game.}
The challenger samples setup and a hidden $\pk_{\mathsf{master}}$, produces honest tags $\{\texttt{Atag}\}$ (for this and other users), and gives the adversary all public data and $\{\texttt{Atag}\}$. The adversary outputs a non-trivial function value $f(\pk_{\mathsf{master}})$. Define
\[
\mathsf{Adv}^{\mathsf{MDA}}_{\mathcal{A}}(\lambda)
=\Pr[\mathcal{A}\ \text{outputs}\ f(\pk_{\mathsf{master}})]-\max_{y}\Pr[y].
\]

\paragraph{Theorem A.1.}
$\mathsf{Adv}^{\mathsf{MDA}}_{\mathcal{A}}(\lambda)\le \mathsf{Adv}^{\mathsf{IND\mbox{-}CPA}}_{\mathsf{ThEnc}} + \mathsf{Adv}^{\mathsf{SEM}}_{\mathsf{PKE\mbox{-}ET}} + \mathsf{negl}(\lambda)$.

\paragraph{Proof.}
Hybrid sequence: 
H$_0$ real; 
H$_1$ simulate $\pi_{\mathsf{link}}$ (zk zero-knowledge) $\Rightarrow$ negl gap; 
H$_2$ replace $\mathsf{UID}=\mathsf{Enc}_{\mathsf{tpk}}(\pk_{\mathsf{master}})$ by $\mathsf{Enc}_{\mathsf{tpk}}(0)$ (threshold IND-CPA); 
H$_3$ replace $\mathsf{CT}^{\mathsf{link}}=\mathsf{ET.Enc}(m)$ by $\mathsf{ET.Enc}(0)$ (PKE-ET semantic security); 
H$_4$ replace $\mathsf{Com}_{\pk}$ by a commitment to $0$ (perfect hiding; zero gap). 
In H$_4$ the distribution is independent of $\pk_{\mathsf{master}}$, so the adversary cannot compute any non-trivial $f(\pk_{\mathsf{master}})$ beyond guessing. Summing gaps yields the bound. \qed

\subsection{Property~\ref{prop:aol}: Auditor-Only Linkability (AOL)}
\label{P2}
\paragraph{Game.}
In challenge bit $b$: if $b{=}1$, two tags come from the \emph{same} user; if $b{=}0$, from \emph{different} users. The adversary receives the two tags \emph{without} the equality-test key and outputs $b'$. Advantage $\mathsf{Adv}^{\mathsf{AOL}}_{\mathcal{A}} = | \Pr[b'{=}b]-\tfrac12 |$.

\paragraph{Theorem A.2.}
$\mathsf{Adv}^{\mathsf{AOL}}_{\mathcal{A}}(\lambda)\le \mathsf{Adv}^{\mathsf{SEM}}_{\mathsf{PKE\mbox{-}ET}}(\lambda)+\mathsf{negl}(\lambda)$.

\paragraph{Proof.}
Simulate $\pi_{\mathsf{link}}$ (zero-knowledge). Replace both challenge $\mathsf{CT}^{\mathsf{link}}$ by encryptions of $0$; the only $b$-dependent difference was whether the plaintexts were equal. Without the equality-test key, PKE-ET semantic security renders the two distributions indistinguishable up to the stated bound. \qed

\paragraph{Auditor correctness.}
An authorized auditor with the equality-test key runs $\mathsf{ET.Test}(\mathsf{CT}^{\mathsf{link}}_i,\mathsf{CT}^{\mathsf{link}}_j)$ which returns $1$ iff the underlying $m_i{=}m_j$. Because $m=H(\mathsf{Com}_{\pk}\Vert \mathsf{ctx}_{\mathsf{GLOBAL}})$ is stable per entity across domains and time, equality holds exactly for same-user pairs.

\subsection{Property~\ref{prop:atuf}: Audit Tag Unforgeability}
\label{P3}
\paragraph{Game.}
The adversary outputs a fresh $\texttt{Atag}^\star$ that passes the audit-chain verifier: valid Axelar threshold signature on provenance fields, valid $\pi_{\mathsf{link}}^\star$, valid execution proof(s) if required, and passes deduplication on $(\mathsf{chain\_id},\mathsf{txid},\mathsf{seq})$.

\paragraph{Theorem A.3.}
\[
\mathsf{Adv}^{\mathsf{ATUF}}_{\mathcal{A}}
\ \le\ 
\mathsf{Adv}^{\mathsf{EUF\mbox{-}CMA}}_{\mathsf{BLS\mbox{-}Thresh}}
+
\mathsf{Adv}^{\mathsf{Sound}}_{\Pi_{\mathsf{SN}}}
+
\mathsf{negl}(\lambda).
\]

\paragraph{Proof.}
If $\texttt{Atag}^\star$ verifies, either (a) the Axelar threshold signature is forged (EUF-CMA break), or (b) some zk proof attests a false statement (soundness break), or (c) it corresponds to a finalized transaction, contradicting freshness given deduplication. Standard reductions give the bound. \qed

\subsection{Property~\ref{prop:rr}: Replay Resistance}
\label{P4}
\paragraph{Game.}
The verifier maintains a set $\mathcal{I}$ of seen tuples $(\mathsf{chain\_id},\mathsf{txid},\mathsf{seq})$. Success means re-accepting an element already in $\mathcal{I}$.

\paragraph{Lemma A.4.}
$\Pr[\text{replay accepted}] = 0$.

\paragraph{Proof.}
Insertion into $\mathcal{I}$ happens \emph{after} acceptance; subsequent submissions of the same tuple are deterministically rejected. Cross-domain replays fail by the $\mathsf{chain\_id}$ binding. This is a functional (non-cryptographic) invariant. \qed

\subsection{Property~\ref{prop:irp}: IRP Privacy and Correctness}
\label{P5}
\paragraph{Privacy game.}
The challenger samples $(X_0,X_1)$ and returns $\mathsf{UID}^\star=\mathsf{Enc}_{\mathsf{tpk}}(X_b)$ for a hidden $b\in\{0,1\}$. The adversary obtains $t'<t$ valid partial decryptions and outputs $b'$. Advantage $| \Pr[b'{=}b]-\tfrac12 |$.

\paragraph{Theorem A.5 (IRP Privacy).}
$\mathsf{Adv}^{\mathsf{IRP\mbox{-}Priv}}_{\mathcal{A}}\le \mathsf{Adv}^{\mathsf{IND\mbox{-}CPA}}_{\mathsf{ThEnc}}$.

\paragraph{Proof.}
A distinguisher against the IRP privacy game yields an IND-CPA adversary for threshold ElGamal: treat the $t'{<}t$ partial decryptions as efficiently simulatable views that leak no information on the plaintext under IND-CPA; simulate missing shares via the public commitments (e.g., Feldman). \qed

\paragraph{Correctness game.}
Given $\mathsf{UID}=\mathsf{Enc}_{\mathsf{tpk}}(X)$, the adversary collects $t$ valid shares but recombination outputs $X'\ne X$ (or fails).

\paragraph{Theorem A.6 (IRP Correctness).}
$\Pr[\text{IRP recombination fails or outputs }X'\ne X]\le \mathsf{negl}(\lambda)$.

\paragraph{Proof.}
With $t$ valid shares, Shamir reconstruction is unique; malformed shares are rejected by verification against public commitments. Hence recombination returns $X$ except with negligible probability. \qed

\subsection{Attack~\ref{attack:post-disclosure} and Post-Disclosure Unforgeability}

\paragraph{Setting.}
After authorized revealing, $\pk_{\mathsf{master}}$ (but not $\sk_{\mathsf{master}}$) becomes public. We show a new valid tag bound to that key cannot be produced.

\paragraph{Game.}
The adversary is given $\pk_{\mathsf{master}}$ and public parameters and outputs a fresh verifying $\texttt{Atag}^\star$ bound (per the circuit) to $\pk_{\mathsf{master}}$.

\paragraph{Theorem A.7.}
\[
\mathsf{Adv}^{\mathsf{PDUF}}_{\mathcal{A}} \le
\mathsf{Adv}^{\mathsf{EUF\mbox{-}CMA}}_{\mathsf{BLS\mbox{-}Thresh}}
+
\mathsf{Adv}^{\mathsf{Sound}}_{\Pi_{\mathsf{SN}}}
+
\mathsf{negl}(\lambda).
\]

\paragraph{Proof.}
A fresh verifying tag requires (i) a valid Axelar threshold signature (EUF-CMA hard), and (ii) a valid $\pi_{\mathsf{link}}$ on a witness consistent with $\pk_{\mathsf{master}}$ (soundness hard). Public knowledge of $\pk_{\mathsf{master}}$ does not enable producing the witness required by the circuit. Hence any successful forger breaks one of the assumptions. \qed

\subsection{Collected Bounds}

\[
\begin{aligned}
\mathsf{Adv}^{\mathsf{MDA}}_{\mathcal{A}} &\le 
\mathsf{Adv}^{\mathsf{IND\mbox{-}CPA}}_{\mathsf{ThEnc}} + 
\mathsf{Adv}^{\mathsf{SEM}}_{\mathsf{PKE\mbox{-}ET}} + \mathsf{negl},\\
\mathsf{Adv}^{\mathsf{AOL}}_{\mathcal{A}} &\le 
\mathsf{Adv}^{\mathsf{SEM}}_{\mathsf{PKE\mbox{-}ET}} + \mathsf{negl},\\
\mathsf{Adv}^{\mathsf{ATUF}}_{\mathcal{A}} &\le 
\mathsf{Adv}^{\mathsf{EUF\mbox{-}CMA}}_{\mathsf{BLS\mbox{-}Thresh}} + 
\mathsf{Adv}^{\mathsf{Sound}}_{\Pi_{\mathsf{SN}}} + \mathsf{negl},\\
\Pr[\text{Replay accepted}] &= 0,\\
\mathsf{Adv}^{\mathsf{IRP\mbox{-}Priv}}_{\mathcal{A}} &\le 
\mathsf{Adv}^{\mathsf{IND\mbox{-}CPA}}_{\mathsf{ThEnc}},\quad
\Pr[\text{IRP incorrect}] \le \mathsf{negl},\\
\mathsf{Adv}^{\mathsf{PDUF}}_{\mathcal{A}} &\le 
\mathsf{Adv}^{\mathsf{EUF\mbox{-}CMA}}_{\mathsf{BLS\mbox{-}Thresh}} + 
\mathsf{Adv}^{\mathsf{Sound}}_{\Pi_{\mathsf{SN}}} + \mathsf{negl}.
\end{aligned}
\]

\section{Measurement Details}
\label{appendix:method}

\subsection{Bootstrap-Based 95\% Confidence Intervals}
\label{appendix:ci}
We report each metric as the mean of $n=10$ independent runs and compute 95\% confidence intervals (CIs) via non-parametric bootstrap with $B=1{,}000$ resamples. Given observations
\[
\mathcal{S}=\{x_1,x_2,\dots,x_n\},
\]
we repeat for each $b\in\{1,\dots,B\}$:
(1) draw a bootstrap sample $\mathcal{S}^{(b)}$ of size $n$ by sampling from $\mathcal{S}$ \emph{with replacement};
(2) compute its mean $\bar{x}^{(b)}$.
Let $\mathcal{M}=\{\bar{x}^{(1)},\dots,\bar{x}^{(B)}\}$ and sort $\mathcal{M}$ ascending. The 95\% CI is the empirical quantile interval
\[
\mathrm{CI}_{95\%}=\big[\,Q_{0.025}(\mathcal{M}),\;Q_{0.975}(\mathcal{M})\,\big].
\]
Plots show the sample mean $\bar{x}=\frac{1}{n}\sum_{i=1}^n x_i$ with error bars equal to $\mathrm{CI}_{95\%}$.

\subsection{Workload Generation}
\label{appendix:workload}
We use two interaction patterns that our prototype supports:
\begin{enumerate}
    \item \textbf{Simple transfer:} a direct cross-chain asset movement (source L1 $\rightarrow$ destination L1) that triggers audit-tag emission and bridging to the audit chain.
    \item \textbf{Escrow settlement:} funds deposited into a neutral escrow on the source chain and released upon satisfaction of a contract predicate (e.g., multi-party acknowledgement), followed by cross-chain completion and audit-tag emission.
\end{enumerate}
Unless otherwise noted, workloads are generated in closed loop: the next transaction is issued after the previous one reaches the configured confirmation depth and the corresponding \Atag{} is persisted on the audit chain.

\subsection{Axelar Confirmation Depth and Provenance Metrics}
\label{appendix:confirm}
We vary the Axelar confirmation depth $d\in\{1,2,4,8\}$, i.e., the number of source-chain block confirmations required before a cross-chain message is accepted for relay. We measure:
\begin{itemize}
    \item \textbf{End-to-end latency:} time from the source-chain transaction commit to the \Atag{} being committed on the audit chain.
    \item \textbf{Replay-safety:} we verify that deduplication on $(\texttt{chain\_id},\texttt{txid})$ (and a monotonic \texttt{seq} if present) rejects replays across reorgs or redundant relays.
\end{itemize}
This isolates the latency~vs.~replay-safety trade-off induced by $d$ without relying on timestamps for correctness.

\subsection{Auditor-Only Linkability (AOL) Evaluation Protocol}
\label{appendix:aol}
We construct ground-truth pairs across chains for equality testing under auditor-only linkability:
\begin{enumerate}
    \item \textbf{Same-entity pairs:} multiple cross-chain transactions produced by the \emph{same} user (same master key) across heterogeneous chains, yielding audit tags with equality on the auditor-side link ciphertexts.
    \item \textbf{Different-entity pairs:} transactions from \emph{distinct} users (different master keys), expected to test unequal.
\end{enumerate}
For each pair $(i,j)$ we run the auditor’s equality test
\[
\ET.\mathsf{Test}(\tk,\CTlink_i,\CTlink_j)\in\{\mathsf{equal},\mathsf{unequal}\}.
\]
We report:
\begin{itemize}
    \item \textbf{AOL correctness:} true-positive rate (TPR) on same-entity pairs and true-negative rate (TNR) on different-entity pairs.
    \item \textbf{Throughput/latency:} number of $\ET.\mathsf{Test}$ operations per second and per-call latency on the auditor machine, using the same bootstrap method (Section~\ref{appendix:ci}) to produce 95\% CIs.
\end{itemize}
Privacy of AOL (i.e., public-side unlinkability without \tk) is covered by the formal proofs and is not re-evaluated empirically.

\subsection{Measurement Conventions}
\label{appendix:conv}
All clocks are taken from the local measurement host unless otherwise specified. For each configuration, we perform $n=10$ independent runs; we randomize transaction nonces and chain-side gas parameters to avoid scheduler artifacts. Reported metrics are the sample mean with 95\% CIs computed via the bootstrap in Section~\ref{appendix:ci}.

\cleardoublepage
\appendix

\cleardoublepage

\bibliographystyle{abbrv}

\end{document}